\documentclass[onecolumn, trackchanges]{aastex631}
\usepackage{enumerate}
\usepackage{import}
\usepackage{xspace}
\usepackage{color}
\usepackage{amsmath}
\usepackage{appendix}
\usepackage{hyperref}

\newcommand{\msun}{\ensuremath{M_\odot}}
\newcommand{\mjup}{\ensuremath{M_J}}
\newcommand{\rjup}{\ensuremath{R_J}}

\newcommand{\kms}{\ensuremath{\mathrm{km\,s^{-1}}}}

\usepackage{etoolbox}

\makeatletter

\patchcmd{\NAT@citex}
  {\@citea\NAT@hyper@{%
     \NAT@nmfmt{\NAT@nm}%
     \hyper@natlinkbreak{\NAT@aysep\NAT@spacechar}{\@citeb\@extra@b@citeb}%
     \NAT@date}}
  {\@citea\NAT@nmfmt{\NAT@nm}%
   \NAT@aysep\NAT@spacechar\NAT@hyper@{\NAT@date}}{}{}

\patchcmd{\NAT@citex}
  {\@citea\NAT@hyper@{%
     \NAT@nmfmt{\NAT@nm}%
     \hyper@natlinkbreak{\NAT@spacechar\NAT@@open\if*#1*\else#1\NAT@spacechar\fi}%
       {\@citeb\@extra@b@citeb}%
     \NAT@date}}
  {\@citea\NAT@nmfmt{\NAT@nm}%
   \NAT@spacechar\NAT@@open\if*#1*\else#1\NAT@spacechar\fi\NAT@hyper@{\NAT@date}}
  {}{}

\makeatother
\received{06/02/2023}
\revised{09/27/2023}
\accepted{10/04/2023}

\submitjournal{AJ}

\shortauthors{Blunt et al.}
\shorttitle{First GRAVITY Observations of HIP 65426 b}

\begin{document}

\pagenumbering{arabic}

\title{First VLTI/GRAVITY Observations of HIP 65426 b: Evidence for a Low or Moderate Orbital Eccentricity}

\author[0000-0002-3199-2888]{Sarah Blunt}
\affiliation{Center for Interdisciplinary Exploration and Research in Astrophysics (CIERA) and Department of Physics and Astronomy, Northwestern University, Evanston, IL 60208, USA}
\affiliation{Department of Astronomy, California Institute of Technology, Pasadena, CA 91125, USA}

\author[0000-0001-6396-8439]{W.~O.~Balmer}
\affiliation{ Department of Physics \& Astronomy, Johns Hopkins University, 3400 N. Charles Street, Baltimore, MD 21218, USA}
\affiliation{ Space Telescope Science Institute, Baltimore, MD 21218, USA}

\author{J.~J.~Wang}
\affiliation{Center for Interdisciplinary Exploration and Research in Astrophysics (CIERA) and Department of Physics and Astronomy, Northwestern University, Evanston, IL 60208, USA}

\author{S.~Lacour}
\affiliation{ LESIA, Observatoire de Paris, PSL, CNRS, Sorbonne Universit\'e, Universit\'e de Paris, 5 place Janssen, 92195 Meudon, France}
\affiliation{ European Southern Observatory, Karl-Schwarzschild-Stra$\ss$ e 2, 85748 Garching, Germany}

\author{S.~Petrus}
\affiliation{Instituto de F\'isica y Astronom\'ia, Facultad de Ciencias, Universidad de Valpara\'iso, Av. Gran Breta\~na 1111, Valpara\'iso, Chile}
\affiliation{N\'ucleo Milenio Formac\'ion Planetaria - NPF, Universidad de Valpara\'iso, Av. Gran Breta\~na 1111, Valpara\'iso, Chile}

\author{G.~Bourdarot}
\affiliation{ Max Planck Institute for extraterrestrial Physics, Giessenbachstra$\ss$ e~1, 85748 Garching, Germany}

\author{J.~Kammerer}
\affiliation{ Space Telescope Science Institute, Baltimore, MD 21218, USA}

\author{N.~Pourr\'e}
\affiliation{ Universit\'e Grenoble Alpes, CNRS, IPAG, 38000 Grenoble, France}

\author{E.~Rickman}
\affiliation{ European Space Agency (ESA), ESA Office, Space Telescope Science Institute, 3700 San Martin Drive, Baltimore, MD 21218, USA}

\author{J.~Shangguan}
\affiliation{ Max Planck Institute for extraterrestrial Physics, Giessenbachstra$\ss$ e~1, 85748 Garching, Germany}

\author{T.~Winterhalder}
\affiliation{ European Southern Observatory, Karl-Schwarzschild-Stra$\ss$ e 2, 85748 Garching, Germany}

\author{R.~Abuter}
\affiliation{ European Southern Observatory, Karl-Schwarzschild-Stra$\ss$ e 2, 85748 Garching, Germany}
\author{A.~Amorim}
\affiliation{ Universidade de Lisboa - Faculdade de Ci\^encias, Campo Grande, 1749-016 Lisboa, Portugal}
\affiliation{ CENTRA - Centro de Astrof\' isica e Gravita\c c\~ao, IST, Universidade de Lisboa, 1049-001 Lisboa, Portugal}
\author{R.~Asensio-Torres}
\affiliation{ Max-Planck-Institut f\"ur Astronomie, K\"onigstuhl 17, 69117 Heidelberg, Germany}
\author{M.~Benisty}
\affiliation{ Universit\'e Grenoble Alpes, CNRS, IPAG, 38000 Grenoble, France}
\author{J.-P.~Berger}
\affiliation{ Universit\'e Grenoble Alpes, CNRS, IPAG, 38000 Grenoble, France}
\author{H.~Beust}
\affiliation{ Universit\'e Grenoble Alpes, CNRS, IPAG, 38000 Grenoble, France}
\author{A.~Boccaletti}
\affiliation{ LESIA, Observatoire de Paris, PSL, CNRS, Sorbonne Universit\'e, Universit\'e de Paris, 5 place Janssen, 92195 Meudon, France}
\author{A.~Bohn}
\affiliation{ Leiden Observatory, Leiden University, P.O. Box 9513, 2300 RA Leiden, The Netherlands}
\author{M.~Bonnefoy}
\affiliation{ Universit\'e Grenoble Alpes, CNRS, IPAG, 38000 Grenoble, France}
\author{H.~Bonnet}
\affiliation{ European Southern Observatory, Karl-Schwarzschild-Stra$\ss$ e 2, 85748 Garching, Germany}
\affiliation{ Universit\'e Grenoble Alpes, CNRS, IPAG, 38000 Grenoble, France}
\author{W.~Brandner}
\affiliation{ Max-Planck-Institut f\"ur Astronomie, K\"onigstuhl 17, 69117 Heidelberg, Germany}
\author{F.~Cantalloube}
\affiliation{ Aix Marseille Univ, CNRS, CNES, LAM, Marseille, France}
\author{P.~Caselli }
\affiliation{ Max Planck Institute for extraterrestrial Physics, Giessenbachstra$\ss$ e~1, 85748 Garching, Germany}
\author{B.~Charnay}
\affiliation{ LESIA, Observatoire de Paris, PSL, CNRS, Sorbonne Universit\'e, Universit\'e de Paris, 5 place Janssen, 92195 Meudon, France}
\author{G.~Chauvin}
\affiliation{ Universit\'e Grenoble Alpes, CNRS, IPAG, 38000 Grenoble, France}
\author{A.~Chavez}
\affiliation{Center for Interdisciplinary Exploration and Research in Astrophysics (CIERA) and Department of Physics and Astronomy, Northwestern University, Evanston, IL 60208, USA}
\author{E.~Choquet}
\affiliation{ Aix Marseille Univ, CNRS, CNES, LAM, Marseille, France}
\author{V.~Christiaens}
\affiliation{ School of Physics and Astronomy, Monash University, Clayton, VIC 3800, Melbourne, Australia}
\author{Y.~Cl\'enet}
\affiliation{ LESIA, Observatoire de Paris, PSL, CNRS, Sorbonne Universit\'e, Universit\'e de Paris, 5 place Janssen, 92195 Meudon, France}
\author{V.~Coud\'e~du~Foresto}
\affiliation{ LESIA, Observatoire de Paris, PSL, CNRS, Sorbonne Universit\'e, Universit\'e de Paris, 5 place Janssen, 92195 Meudon, France}
\author{A.~Cridland}
\affiliation{ Leiden Observatory, Leiden University, P.O. Box 9513, 2300 RA Leiden, The Netherlands}
\author{R.~Dembet}
\affiliation{ LESIA, Observatoire de Paris, PSL, CNRS, Sorbonne Universit\'e, Universit\'e de Paris, 5 place Janssen, 92195 Meudon, France}
\author{A.~Drescher}
\affiliation{ Max Planck Institute for extraterrestrial Physics, Giessenbachstra$\ss$ e~1, 85748 Garching, Germany}
\author{G.~Duvert}
\affiliation{ Universit\'e Grenoble Alpes, CNRS, IPAG, 38000 Grenoble, France}
\author{A.~Eckart}
\affiliation{Institute of Physics, University of Cologne, Z\"ulpicher Stra$\ss$ e 77, 50937 Cologne, Germany}
\affiliation{ Max Planck Institute for Radio Astronomy, Auf dem H\"ugel 69, 53121 Bonn, Germany}
\author{F.~Eisenhauer}
\affiliation{ Max Planck Institute for extraterrestrial Physics, Giessenbachstra$\ss$ e~1, 85748 Garching, Germany}
\author{H.~Feuchtgruber}
\affiliation{ Max Planck Institute for extraterrestrial Physics, Giessenbachstra$\ss$ e~1, 85748 Garching, Germany}
\author{P.~Garcia}
\affiliation{ CENTRA - Centro de Astrof\' isica e Gravita\c c\~ao, IST, Universidade de Lisboa, 1049-001 Lisboa, Portugal}
\affiliation{ Universidade do Porto, Faculdade de Engenharia, Rua Dr. Roberto Frias, 4200-465 Porto, Portugal}
\author{R.~Garcia~Lopez}
\affiliation{ School of Physics, University College Dublin, Belfield, Dublin 4, Ireland}
\affiliation{ Max-Planck-Institut f\"ur Astronomie, K\"onigstuhl 17, 69117 Heidelberg, Germany}
\author{E.~Gendron}
\affiliation{ LESIA, Observatoire de Paris, PSL, CNRS, Sorbonne Universit\'e, Universit\'e de Paris, 5 place Janssen, 92195 Meudon, France}
\author{R.~Genzel}
\affiliation{ Max Planck Institute for extraterrestrial Physics, Giessenbachstra$\ss$ e~1, 85748 Garching, Germany}
\author{S.~Gillessen}
\affiliation{ Max Planck Institute for extraterrestrial Physics, Giessenbachstra$\ss$ e~1, 85748 Garching, Germany}
\author{J.~H.~Girard}
\affiliation{ Space Telescope Science Institute, Baltimore, MD 21218, USA}
\author{X.~Haubois}
\affiliation{ European Southern Observatory, Casilla 19001, Santiago 19, Chile}
\author{G.~Hei$\ss$el}
\affiliation{Advanced Concepts Team, European Space Agency, TEC-SF, ESTEC, Keplerlaan 1, 2201 AZ Noordwijk, The Netherlands}
\affiliation{ LESIA, Observatoire de Paris, PSL, CNRS, Sorbonne Universit\'e, Universit\'e de Paris, 5 place Janssen, 92195 Meudon, France}
\author{Th.~Henning}
\affiliation{ Max-Planck-Institut f\"ur Astronomie, K\"onigstuhl 17, 69117 Heidelberg, Germany}
\author{S.~Hinkley}
\affiliation{ University of Exeter, Physics Building, Stocker Road, Exeter EX4 4QL, United Kingdom}
\author{S.~Hippler}
\affiliation{ Max-Planck-Institut f\"ur Astronomie, K\"onigstuhl 17, 69117 Heidelberg, Germany}
\author{M.~Horrobin}
\affiliation{Institute of Physics, University of Cologne, Z\"ulpicher Stra$\ss$ e 77, 50937 Cologne, Germany}
\author{M.~Houll\'e}
\affiliation{ Universit\'e C\^ote d’Azur, Observatoire de la C\^ote d’Azur, CNRS, Laboratoire Lagrange, Nice, France}
\author{Z.~Hubert}
\affiliation{ Universit\'e Grenoble Alpes, CNRS, IPAG, 38000 Grenoble, France}
\author{L.~Jocou}
\affiliation{ Universit\'e Grenoble Alpes, CNRS, IPAG, 38000 Grenoble, France}
\author{M.~Keppler}
\affiliation{ Max-Planck-Institut f\"ur Astronomie, K\"onigstuhl 17, 69117 Heidelberg, Germany}
\author{P.~Kervella}
\affiliation{LESIA, Observatoire de Paris, Universit\'e PSL, CNRS, Sorbonne Universit\'e, Universit\'e Paris Cit\'e, 5 place Jules Janssen, 92195 Meudon, France}
\author{L.~Kreidberg}
\affiliation{ Max-Planck-Institut f\"ur Astronomie, K\"onigstuhl 17, 69117 Heidelberg, Germany}
\author{A.-M.~Lagrange}
\affiliation{ Universit\'e Grenoble Alpes, CNRS, IPAG, 38000 Grenoble, France}
\affiliation{ LESIA, Observatoire de Paris, PSL, CNRS, Sorbonne Universit\'e, Universit\'e de Paris, 5 place Janssen, 92195 Meudon, France}
\author{V.~Lapeyr\`ere}
\affiliation{ LESIA, Observatoire de Paris, PSL, CNRS, Sorbonne Universit\'e, Universit\'e de Paris, 5 place Janssen, 92195 Meudon, France}
\author{J.-B.~Le~Bouquin}
\affiliation{ Universit\'e Grenoble Alpes, CNRS, IPAG, 38000 Grenoble, France}
\author{P.~L\'ena}
\affiliation{ LESIA, Observatoire de Paris, PSL, CNRS, Sorbonne Universit\'e, Universit\'e de Paris, 5 place Janssen, 92195 Meudon, France}
\author{D.~Lutz}
\affiliation{ Max Planck Institute for extraterrestrial Physics, Giessenbachstra$\ss$ e~1, 85748 Garching, Germany}
\author{A.-L.~Maire}
\affiliation{ Universit\'e Grenoble Alpes, CNRS, IPAG, 38000 Grenoble, France}
\author{F.~Mang}
\affiliation{ Max Planck Institute for extraterrestrial Physics, Giessenbachstra$\ss$ e~1, 85748 Garching, Germany}
\author[0000-0002-2919-7500]{G.-D.~Marleau}
\affiliation{Fakult\"at f\"{u}r Physik, Universit\"{a}t Duisburg-Essen, Lotharstraße 1, 47057 Duisburg, Germany}
\affiliation{ Instit\"{u}t f\"{u}r Astronomie und Astrophysik, Universit\"{a}t T\"{u}bingen, Auf der Morgenstelle 10, 72076 T\"{u}bingen, Germany}
\affiliation{Physikalisches Institut, Universit\"{a}t Bern, Gesellschaftsstr.\ 6, 3012 Bern, Switzerland}
\affiliation{ Max-Planck-Institut f\"ur Astronomie, K\"onigstuhl 17, 69117 Heidelberg, Germany}
\author{A.~M\'erand}
\affiliation{ European Southern Observatory, Karl-Schwarzschild-Stra$\ss$ e 2, 85748 Garching, Germany}
\author{P.~Molli\`ere}
\affiliation{ Max-Planck-Institut f\"ur Astronomie, K\"onigstuhl 17, 69117 Heidelberg, Germany}
\author{J.~D.~Monnier}
\affiliation{ Astronomy Department, University of Michigan, Ann Arbor, MI 48109 USA}
\author[0000-0002-1013-2811]{C.~Mordasini}
\affiliation{Physikalisches Institut, Universit\"{a}t Bern, Gesellschaftsstr.\ 6, 3012 Bern, Switzerland}
\author{D.~Mouillet}
\affiliation{ Universit\'e Grenoble Alpes, CNRS, IPAG, 38000 Grenoble, France}
\author{E.~Nasedkin}
\affiliation{ Max-Planck-Institut f\"ur Astronomie, K\"onigstuhl 17, 69117 Heidelberg, Germany}
\author{M.~Nowak}
\affiliation{ Institute of Astronomy, University of Cambridge, Madingley Road, Cambridge CB3 0HA, United Kingdom}
\author{T.~Ott}
\affiliation{ Max Planck Institute for extraterrestrial Physics, Giessenbachstra$\ss$ e~1, 85748 Garching, Germany}
\author{G.~P.~P.~L.~Otten}
\affiliation{Academia Sinica, Institute of Astronomy and Astrophysics, 11F Astronomy-Mathematics Building, NTU/AS campus, No. 1, Section 4, Roosevelt Rd., Taipei 10617, Taiwan}
\author{C.~Paladini}
\affiliation{ European Southern Observatory, Casilla 19001, Santiago 19, Chile}
\author{T.~Paumard}
\affiliation{ LESIA, Observatoire de Paris, PSL, CNRS, Sorbonne Universit\'e, Universit\'e de Paris, 5 place Janssen, 92195 Meudon, France}
\author{K.~Perraut}
\affiliation{ Universit\'e Grenoble Alpes, CNRS, IPAG, 38000 Grenoble, France}
\author{G.~Perrin}
\affiliation{ LESIA, Observatoire de Paris, PSL, CNRS, Sorbonne Universit\'e, Universit\'e de Paris, 5 place Janssen, 92195 Meudon, France}
\author{O.~Pfuhl}
\affiliation{ European Southern Observatory, Karl-Schwarzschild-Stra$\ss$ e 2, 85748 Garching, Germany}
\author{L.~Pueyo}
\affiliation{ Space Telescope Science Institute, Baltimore, MD 21218, USA}
\author{J.~Rameau}
\affiliation{ Universit\'e Grenoble Alpes, CNRS, IPAG, 38000 Grenoble, France}
\author{L.~Rodet}
\affiliation{ Center for Astrophysics and Planetary Science, Department of Astronomy, Cornell University, Ithaca, NY 14853, USA}
\author{Z.~Rustamkulov }
\affiliation{ Department of Earth \& Planetary Sciences, Johns Hopkins University, Baltimore, MD, USA}

\author{T.~Shimizu }
\affiliation{ Max Planck Institute for extraterrestrial Physics, Giessenbachstra$\ss$ e~1, 85748 Garching, Germany}
\author{D.~Sing }
\affiliation{ Department of Physics \& Astronomy, Johns Hopkins University, 3400 N. Charles Street, Baltimore, MD 21218, USA}
\affiliation{ Department of Earth \& Planetary Sciences, Johns Hopkins University, Baltimore, MD, USA}
\author{T.~Stolker}
\affiliation{ Leiden Observatory, Leiden University, P.O. Box 9513, 2300 RA Leiden, The Netherlands}
\author{C.~Straubmeier}
\affiliation{Institute of Physics, University of Cologne, Z\"ulpicher Stra$\ss$ e 77, 50937 Cologne, Germany}
\author{E.~Sturm}
\affiliation{ Max Planck Institute for extraterrestrial Physics, Giessenbachstra$\ss$ e~1, 85748 Garching, Germany}
\author{L.~J.~Tacconi}
\affiliation{ Max Planck Institute for extraterrestrial Physics, Giessenbachstra$\ss$ e~1, 85748 Garching, Germany}
\author{E.F.~van~Dishoeck}
\affiliation{ Leiden Observatory, Leiden University, P.O. Box 9513, 2300 RA Leiden, The Netherlands}
\affiliation{ Max Planck Institute for extraterrestrial Physics, Giessenbachstra$\ss$ e~1, 85748 Garching, Germany}
\author{A.~Vigan}
\affiliation{ Aix Marseille Univ, CNRS, CNES, LAM, Marseille, France}
\author{F.~Vincent}
\affiliation{ LESIA, Observatoire de Paris, PSL, CNRS, Sorbonne Universit\'e, Universit\'e de Paris, 5 place Janssen, 92195 Meudon, France}
\author{K.~Ward-Duong}
\affiliation{ Department of Astronomy, Smith College, Northampton MA 01063 USA}
\author{F.~Widmann}
\affiliation{ Max Planck Institute for extraterrestrial Physics, Giessenbachstra$\ss$ e~1, 85748 Garching, Germany}
\author{E.~Wieprecht}
\affiliation{ Max Planck Institute for extraterrestrial Physics, Giessenbachstra$\ss$ e~1, 85748 Garching, Germany}
\author{E.~Wiezorrek}
\affiliation{ Max Planck Institute for extraterrestrial Physics, Giessenbachstra$\ss$ e~1, 85748 Garching, Germany}
\author{J.~Woillez}
\affiliation{ European Southern Observatory, Karl-Schwarzschild-Stra$\ss$ e 2, 85748 Garching, Germany}
\author{S.~Yazici}
\affiliation{ Max Planck Institute for extraterrestrial Physics, Giessenbachstra$\ss$ e~1, 85748 Garching, Germany}
\author{A.~Young}
\affiliation{ Max Planck Institute for extraterrestrial Physics, Giessenbachstra$\ss$ e~1, 85748 Garching, Germany}

\author{The exoGRAVITY collaboration}


\begin{abstract}

Giant exoplanets have been directly imaged over orders of magnitude of orbital separations, prompting theoretical and observational investigations of their formation pathways. In this paper, we present new VLTI/GRAVITY astrometric data of HIP 65426 b, a cold, giant exoplanet which is a particular challenge for most formation theories at a projected separation of 92 au from its primary. Leveraging GRAVITY's astrometric precision, we present an updated eccentricity posterior that disfavors large eccentricities. The eccentricity posterior is still prior-dependent, and we extensively interpret and discuss the limits of the posterior constraints presented here. We also perform updated spectral comparisons with self-consistent forward-modeled spectra, finding a best fit \texttt{ExoREM} model with solar metallicity and C/O=0.6. An important caveat is that it is difficult to estimate robust errors on these values, which are subject to interpolation errors as well as potentially missing model physics. Taken together, the orbital and atmospheric constraints paint a preliminary picture of formation inconsistent with scattering after disk dispersal. Further work is needed to validate this interpretation. Analysis code used to perform this work is available at https://github.com/sblunt/hip65426. 

\end{abstract}

\section{Introduction}

The existence of cold Jupiters (CJs) at large separations (10s of au) is a formation puzzle. Direct imaging surveys have revealed that these planets are intrinsically rare, with an occurrence rate of about 1\% \citep{Bowler:2018a,Vigan:2021}, but given that the nominal core formation timescale at typical CJ separations is much longer than the disk dispersal timescale \citep{Armitage:2020a}, we would not expect CJs to exist at all, assuming in-situ core accretion. Direct gravitational collapse within a disk is an alternative to core accretion, but with its own issues, particularly very fast migration after formation (\citealp{Nayakshin:2017a}, \citealp{Vorobyov:2018a}). The CJ population is quite diverse, with masses spanning an order of magnitude and separations spanning several \citep{Bowler:2016a}, so it is also possible that multiple formation mechanisms are at play.

The eccentricities of CJs are a useful tracer of formation history, both at the population level \citep{Bowler:2020a}, and for individual systems. \citet{Marleau:2019a} conducted a detailed investigation of possible formation scenarios via core accretion for the $\sim$14 Myr CJ HIP 65426 b (Table \ref{tab:hip654}), following on the work of \citet{ColemanNelson:2016a,ColemanNelson:2016b}. At a projected separation of 92 au \citep{Chauvin:2017a}, this object is significantly farther from its star than, for example, the famous HR 8799 planets (at 16, 26, 41, and 72 au, \citealp{Wang:2018a}), and therefore even more of a challenge for formation via core accretion, as core formation timescale increases with separation in the widely-separated regime. \citet{Marleau:2019a} derived distributions over the initial mass and luminosity of HIP 65426 b under several assumptions of post-formation entropy, then coupled these initial conditions with N-body simulations to investigate which models could match the present-day conditions of the planet. They varied the initial number of planets in the system, and included prescriptions for type I and II migration in a protoplanetary disk. Through a suite of such simulations, they found two families of explanations for HIP 65426 b's current separation and luminosity. The first is core formation at close separations, followed by outward scattering and subsequent runaway gas accretion at the present-day location. The slower timescale of type II migration and subsequent disk dispersal allows the planet to remain in place after scattering and damps the post-scattering eccentricity. The second scenario is similar, except runaway accretion and subsequent disk dispersal occur \textit{before} scattering. Under this scenario, most simulations resulted in a high ($>$0.5) present-day eccentricity for HIP 65426 b. A third scenario the authors mentioned but did not investigate in detail is the prospect of in-situ formation, invoking a more rapid core formation process than typically assumed, such as pebble accretion (\citealp{Lambrechts:2014a}, \citealp{Rosenthal:2018a}). This scenario would presumably result in a circular orbit. 

In summary, \citet{Marleau:2019a} propose three statistically distinct formation pathways via core accretion for HIP 65426 b: 1)~in-situ formation, resulting in a circular orbit, 2)~scattering before disk dispersal, resulting in a low-to-moderate eccentricity orbit, and 3)~scattering after disk dispersal, resulting in a high eccentricity orbit. Both scenarios 2 and 3 are often accompanied by an inner giant planet. An important caveat is that these are not hard-and-fast distinctions; an eccentricity of 0, even with no model uncertainty, would not unequivocally rule out two of the three scenarios. However, such a measurement will allow us to assign statistical probabilities to each formation scenario by comparing with the population synthesis outputs of studies like \citet{Marleau:2019a}. Precise measurements of eccentricity will also better constrain the population-level eccentricity distribution of CJs, which is currently still prior-dependent \citep{Nagpal:2023a}, allowing us to model the formation mechanisms responsible for the population as a whole.

HIP 65426 b has been astrometrically monitored since its discovery by \citet{Chauvin:2017a}, and previous papers report an essentially unconstrained eccentricity posterior that reproduces the prior (\citealp{Chauvin:2017a}, \citealp{Cheetham:2019a}, \citealp{Carter:2022a}). In this paper, we update the orbit model of HIP 65426 b using new astrometry from the optical interferometer VLTI/GRAVITY ($\sim$50x more precise than previous astrometry) as part of the ExoGRAVITY project \citep{Lacour:2020a}.

Spectral and photometric measurements of HIP 65426 b have also been used to infer the planet's atmospheric properties. \citet{Petrus:2021a} recovered a bimodal posterior by comparing the HIP 65426 b measurements with \texttt{BT-Settl CIFIST} models, with one small radius ($\sim$1\rjup{}) peak and one larger radius peak ($\sim$1.2\rjup{}). Their \texttt{Exo-REM} comparisons yielded broader posteriors encompassing both of these possibilities, as \texttt{Exo-REM} predictions were only available for the K-band at the time, and they could not compare with all available data. These yielded  T$_{\mathrm{eff}}$=$1560\pm100 K$, $\log{g}<4.40$, a slightly super-solar metallicity of $0.05^{+0.24}_{-0.22}$, and an upper limit on the C/O ratio ($\leq0.55$). \citet{Carter:2022a} subsequently published 7 long-wavelength photometric datapoints of HIP 65426 b using the NIRCAM and MIRI instruments on JWST, and used them to update the empirical bolometric luminosity and of the planet to $\log{\mathrm{L/L}_{\odot}}$ = (-4.31, -4.14). Together with hot-start cooling models, they used this luminosity estimate to infer a planetary mass, radius, and effective temperature (given in Table \ref{tab:hip654}). They found that an independent comparison to the \texttt{BT-Settl CIFIST} grid yielded a good fit, but resulted in an unphysically small planetary radius of $1.06\pm0.05$ \rjup{} (consistent with the smaller radius mode of \citealp{Petrus:2021a}). 

This paper is structured as follows: in Section \ref{sec:data}, we present our new GRAVITY data, including three new astrometric epochs and a new K-band spectrum. In Section \ref{sec:orbit}, we present our updated orbital solution and discuss the significance of our eccentricity measurement in detail. In Section \ref{sec:spectrum}, we compare all existing photometry and spectra measurements of HIP 65426 with self-consistent model spectra to update our understanding of the planet's atmospheric properties, finding results that are consistent with but more precise than previous work, keeping in mind that systematic uncertainties in these fits are likely underestimated. Finally, in Section \ref{sec:discuss} we discuss the implications of our orbital and atmospheric inferences and call for additional observational and theoretical work.

\begin{deluxetable*}{ccc}
    \tablecaption{Relevant physical properties of HIP 65426 A and b\label{tab:hip654}}
    \tablehead{Quantity & Value & Source}
\startdata
Stellar spectral type & A2V & \citet{Carter:2022a} \\
Stellar mass & $1.96\pm 0.04~\msun$ & \cite{Chauvin:2017a} \\
Moving group membership & Lower Centaurus-Crux, $14\pm4$ Myr & \citet{Gagne:2018a} \\
Stellar parallax & $9.30\pm0.03$ mas & Gaia DR3 (\citealt{Gaia:2016a}, \\
&&\citealt{Gaia:2022a})\\
Hot-start planetary mass estimate & $7.1\pm1.2$ \mjup{} & \citet{Carter:2022a}\\
Hot-start planetary radius estimate & $1.44\pm0.03$ \rjup{} & \citet{Carter:2022a}\\
Hot-start planetary T$_{\mathrm{eff}}$ & $1283^{+25}_{-31}$ K & \citet{Carter:2022a}\\
\enddata
\end{deluxetable*}

\section{Data}
\label{sec:data}

\subsection{GRAVITY Data}

\begin{deluxetable*}{ccccccccr}
\tablewidth{\textwidth}
\tablecaption{ Observing log. NEXP, NDIT, and DIT denote the number of exposures, the number of detector integrations per exposure, and the detector integration time, respectively. $\tau_0$ is the atmospheric coherence time during each exposure. The fiber pointing is the placement of the science fiber relative to the fringe tracking fiber (which is placed on the central star). HD\,91881 and HD\,73900 are two binary systems used for phase referencing.
\label{tab:obslog}}
\tablehead{Date & \multicolumn{2}{c}{UT time} & Target & NEXP/NDIT/DIT & Airmass & $\tau_0$ & Seeing & Fiber pointing  \\
& Start & End &  &  & & & & $\Delta$RA/$\Delta$DEC}
\startdata
\hline\hline
2021-01-07 & 06:22:05 & 06:48:54 & HD\,91881 & 8 / 64 / 1\,s & 1.05--1.10 & 6.6--8.0~ms & 0.51--0.65$^{\prime\prime}$ & -1108 / 710 mas  \\
2021-01-07 & 06:59:10 & 07:57:00 & HIP\,65426\,b & 4 / 8 / 100\,s & 1.38--1.64 & 5.0--6.6~ms & 0.61--1.19$^{\prime\prime}$ & 416 / -704 mas  \\
2021-01-07 & 08:03:17 & 08:06:18 & HIP\,65426\,A & 2 / 64 / 1\,s & 1.35--1.36 & 4.6--5.4~ms & 1.05--1.23$^{\prime\prime}$ & 0 / 0 mas  \\
\hline\hline
2022-01-23 & 05:16:56 & 05:30:01 & HD\,73900 & 6 / 64 / 1\,s & 1.05--1.10 & 13.2--20.4~ms & 0.41--0.57$^{\prime\prime}$ & -825 / -455 mas  \\
2022-01-23 & 05:42:05 & 05:55:37 & HD\,91881 & 6 / 64 / 1\,s & 1.38--1.64 & 6.4--13.8~ms & 0.30--0.67$^{\prime\prime}$ & -1108 / 710 mas  \\
2022-01-23 & 06:08:31 & 06:51:52 & HIP\,65426\,b & 6 / 4 / 100\,s & 1.35--1.36 & 5.4--13.9~ms & 0.54--0.72$^{\prime\prime}$ & 418 / -699 mas  \\
\hline\hline
2023-05-07 & 04:18:13 & 05:30:00 & HIP\,65426\,b & 6 / 4 / 100\,s & 1.16--1.29 & 1.6--2.0~ms & 1.34--1.90$^{\prime\prime}$ & 419 / -696 mas  \\
2023-05-07 & 05:47:00 & 05:55:52 & HD\,123227 & 4 / 96 / 0.3\,s & 1.20--1.22 & 1.8--2.1~ms & 1.26--1.48$^{\prime\prime}$ & -404/ 889 mas  \\
\hline\hline
\enddata
\end{deluxetable*}

\par We observed HIP~65426~b on the 7th of January, 2021, the 23rd of January, 2022, and the 7th of May, 2023 as part of the ExoGRAVITY Large Program \citep[ESO Program ID 1104.C-0651,][]{Lacour:2020a}. 
We used the European Southern Observatory (ESO) Very Large Telescope Interferometer (VLTI)'s four 8.2m Unit Telescopes (UTs) and the GRAVITY instrument \citep{GravityCollaboration:2017a}.
The observations primarily used the ``off-axis dual field'' mode, in which a roof mirror is used to split the telescope field into two. The star light goes to the 
fringe tracker to correct for atmospheric perturbations \citep{Lacour:2019a}. The exoplanetary light goes to the spectrograph configured with the medium resolution grism (R=500). Phase referencing of the metrology is obtained by swapping on a binary just before the observation. On 2021-01-07 we used the binary system HD\,91881, on 2022-01-23 we used two binary systems: HD\,73900 and HD\,91881, and on 2023-05-07 we used HD\,123227. During the night of the 2021-01-07, we also performed an observation ``on-axis single-field'' observations of the host star to calibrate the spectrum of the planet for that night.
 The second and third nights of observations do not have an on-axis calibrator, and so we do not consider the spectrum of the planet from those observations. The observing log, presented in Table \ref{tab:obslog}, records the length of the observations, the number of files recorded, and the atmospheric conditions. The placement of the science fiber was based on preliminary orbit predictions fit to the available relative astrometry at the time using the \texttt{whereistheplanet}\footnote{\url{http://www.whereistheplanet.com}} software \citep{whereistheplanet}, and resulted in an efficient coupling of the planet flux into the science fiber.

\par For the 2021-01-07 epoch, we calculated the complex visibilites of the host and the companion using the Public Release 1.5.0 (1 July 2021\footnote{\url{https://www.eso.org/sci/software/pipelines/gravity/}}) of the ESO GRAVITY pipeline \citep{Lapeyrere:2014a}. The observations were phase-referenced with the metrology system using observations of the binary calibrator. We decontaminated the flux of the planet due to the host using a custom python pipeline \citep[see Appendix A, ][]{GravityCollaboration:2020a}, which treats the contamination as a polynomial dependent on time and baseline; a polynomial of fourth order was used for stellar light suppression.

\par
\begin{figure*}
    \centering
    \includegraphics[width=0.7\textwidth]{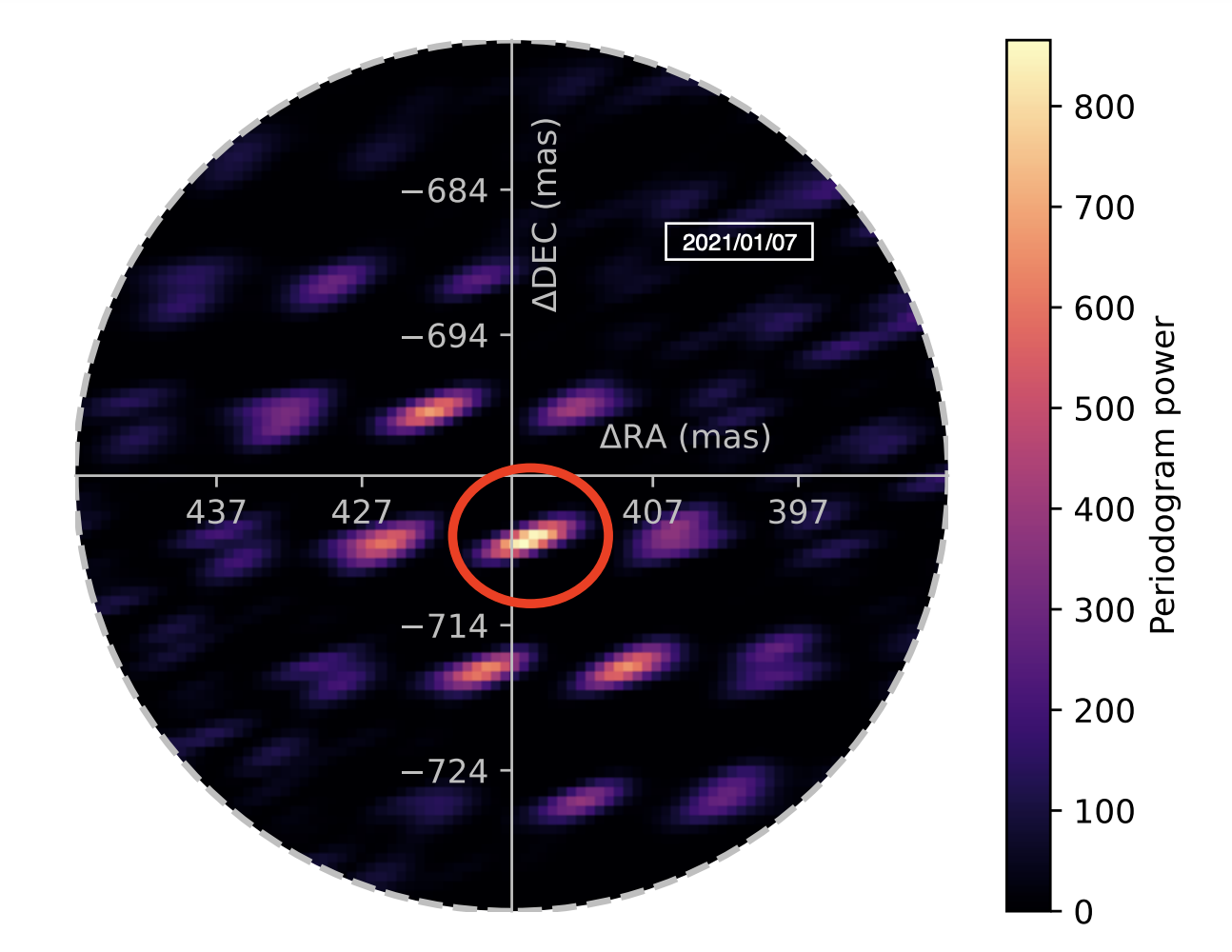}
    \includegraphics[width=0.7\textwidth, trim={0 0 0.5cm 0}]{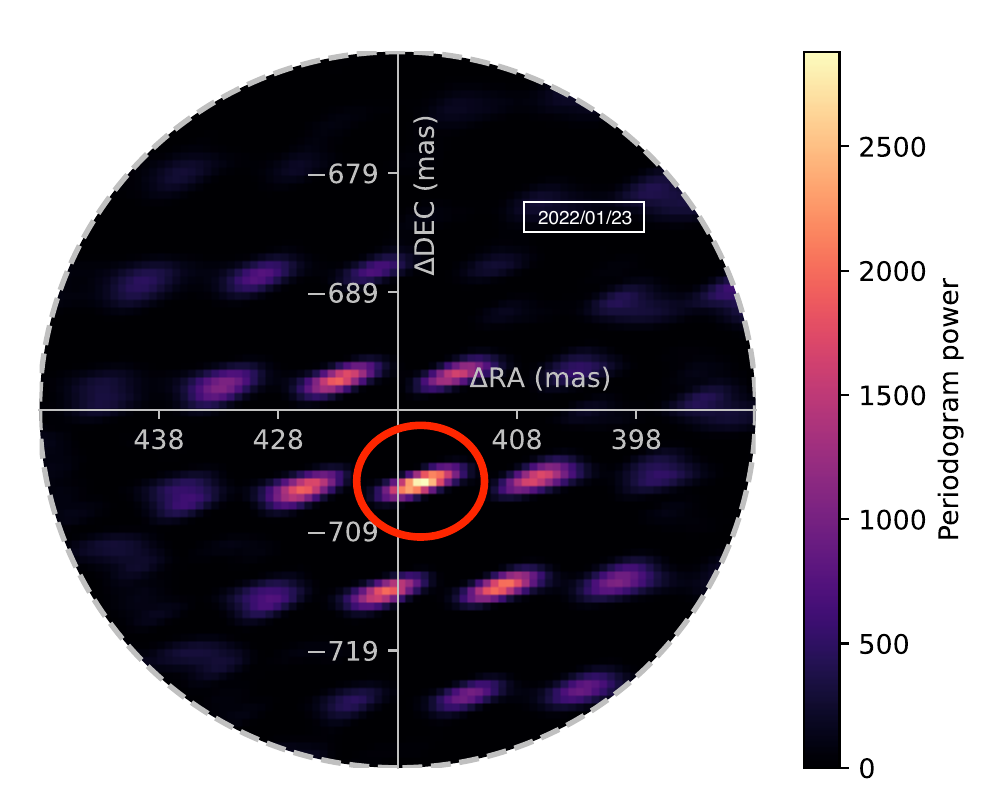}
    \caption{Detections of HIP~65426~b with VLTI/GRAVITY in epochs 1 and 2. Both periodogram power maps visualize the $\chi^2$ fit to the interferometric observables assuming a point source, after removing the contribution of the star using a 4th-order polynomial. The outer dashed grey circle indicates the effective fiber field of view (60~mas in diameter), and the red circles indicate the most probable planet position at each epoch. The planet is detected at high confidence in both epochs (periodogram power $> 500$).}
    \label{fig:chi2map}
\end{figure*}

\begin{figure*}
    \centering
    \includegraphics[width=0.7\textwidth]{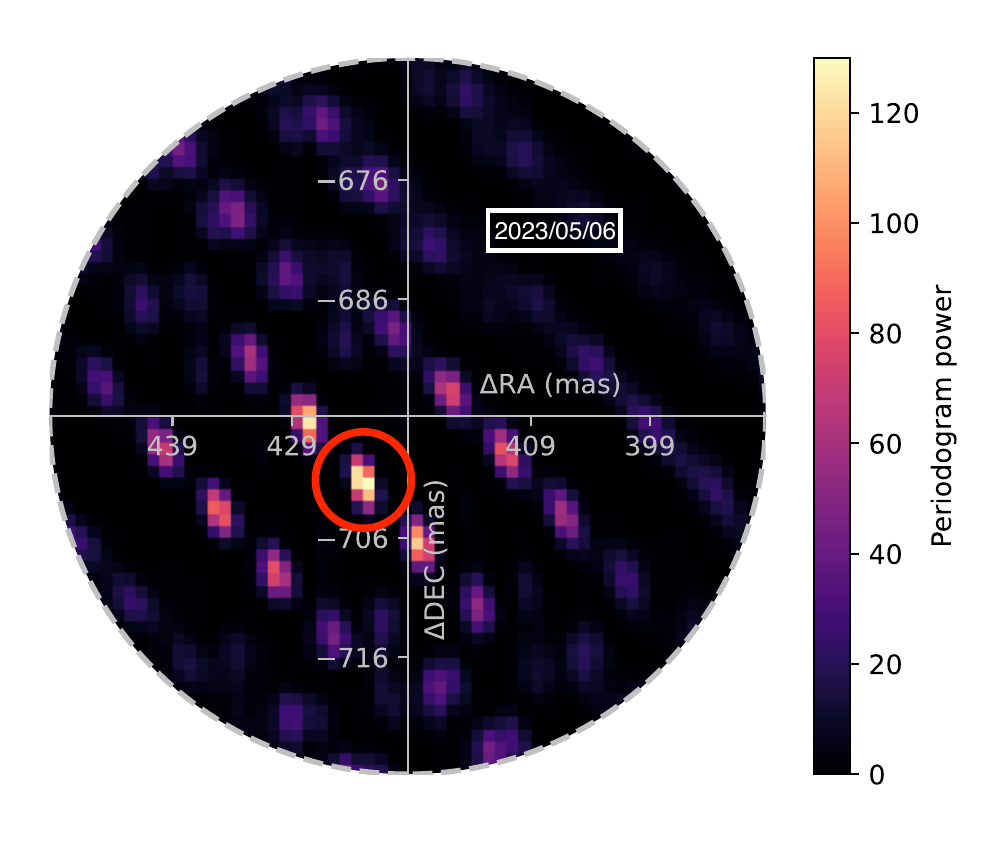}
    \caption{Detection of HIP~65426~b with VLTI/GRAVITY in epoch 3. See Figure \ref{fig:chi2map}.}
    \label{fig:chi2map2}
\end{figure*}

 We obtained the astrometric position of the planet relative to the star at each epoch by analysing the phase of the ratio of coherent fluxes, computing a periodogram power map over the fiber's field-of-view (Figures \ref{fig:chi2map} and \ref{fig:chi2map2}). The mean astrometric position is taken to be the minimum $\chi^2$ value of this map. We estimated the uncertainty on each astrometric measurement using the scatter of mean astrometric values between individual exposures. These new astrometric datapoints are provided in Table \ref{tab:astrom}. 

\par
\begin{figure*}
    \centering
    \includegraphics[width=\textwidth]{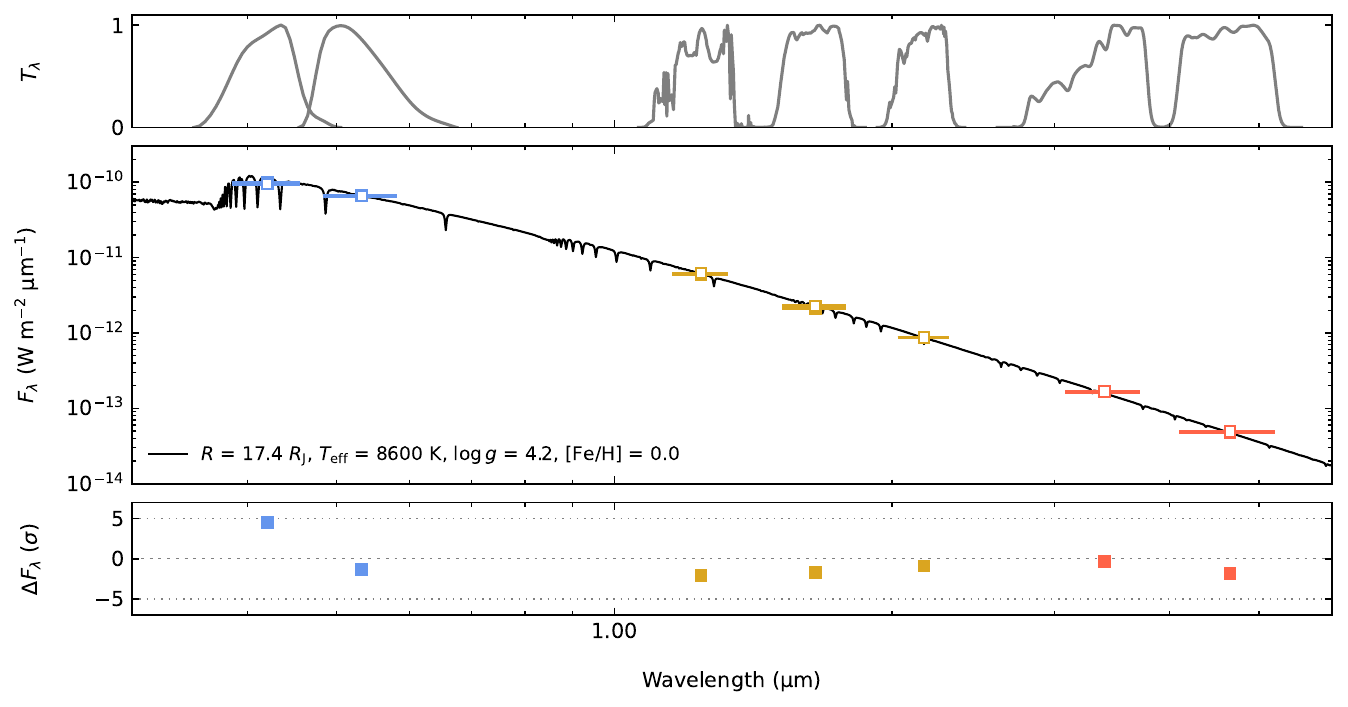}
    \caption{Best-fit SED model of HIP 65426 A used to convert our \textbf{Top:} transmission functions of photometric bands (each plotted as a data point with errors immediately beneath). \textbf{Middle:} MAP BT-NextGen model and photometric data from TYCHO, 2MASS, and WISE. \textbf{Bottom:} MAP model residuals.}
    \label{fig:Ased}
\end{figure*}

 We then extracted the ratio of the coherent flux between the two sources at the location of the companion, generating a contrast spectrum from 2 to 2.5 microns. We only extract a spectrum for the observations from 2021-01-07, where we observed the star in on-axis mode directly after observing the planet in off-axis mode. We converted this contrast spectrum into a flux calibrated spectrum by multiplying it by a BT-NextGen \citep{Allard:2012a} spectrum fit to archive photometry of the host. We used a BT-NextGen model with $\mathrm{T_{eff}}=8600\,\mathrm{K}$ \citep{Carter:2022a} and $\mathrm{log(g)}=4.2$ \citep{Bochanski2018}, and scaled the model to fit the photometric measurements from Tycho2 \citep{Hog2000}, 2MASS \citep{Cutri2003}, and WISE \citep{Wright2010} using \texttt{species} \citep{Stolker:2020a}. The best-fit model is shown in Figure \ref{fig:Ased}. We noted a tension between the absolute flux of the resultant spectrum and published SPHERE K-band photometry \citep{Chauvin:2017a} and SINFONI spectrum \citep{Petrus:2021a}.
For our 2021-01-07 observation, $F_{K2} = 1.40\times10^{-16}~\mathrm{W/m^2/\mu m}$ vs the SPHERE value $F_{K2} = 7.21\times10^{-17}~\mathrm{W/m^2/\mu m}$. This is likely due to two factors: the low integration time on the star ($<3$ minutes) and the degradation/variability of observing conditions between the observation of the planet and the star. Since SPHERE and SINFONI have more robust measurements of the stellar flux during the course of their observations (from satellite spots for SPHERE, and direct measurements of the stellar flux for SINFONI), we opted to integrate the GRAVITY spectrum over the Paranal/IRDIS\_D\_K12\_2 filter and scale the GRAVITY spectrum by $\times0.52$ to match the SPHERE K2 photometry. The existing SPHERE photometry matches the existing SINFONI spectrum (described more and compared with our new GRAVITY spectrum in Section \ref{sec:spectrum}) without any scaling factor applied.

The fluxes, uncertainties, and inter-channel flux covariances are available as a machine-readable table published along with this paper.

\begin{deluxetable*}{cccccc}
    \tablewidth{\textwidth}
    \tablecaption{New relative astrometry of HIP 65426 b presented in this paper. $\sigma_{\Delta \rm R.A.}$ and $\sigma_{\Delta \rm Dec.}$ denote the uncertainties in astrometric position, and $\rho_{\rm \Delta R.A., \Delta Decl.}$ denotes the correlation between the $\sigma_{\Delta \rm R.A.}$ and $\sigma_{\Delta \rm Dec.}$ measurements.
    \label{tab:astrom}}
    \tablehead{ Date & $\Delta$R.A. & $\sigma_{\Delta \rm R.A.}$ & $\Delta$Decl. & $\sigma_{\Delta \rm Decl.}$ & $\rho_{\rm \Delta R.A., \Delta Decl.}$ \\ $[\mathrm{JD} - 2400000.5]$ & [mas] & [mas] & [mas] & [mas] & \\}
\startdata
59221.312 & 415.613 & 0.107 & $-708.133$ & 0.073 & $-0.095$ \\
59602.271 & 416.269 & 0.035 & $-705.051$ & 0.035 & $-0.224$ \\
60071.208 & 416.980 & 0.281 & $-701.373$ & 0.248 & 0.954 \\
\enddata
\end{deluxetable*}

\begin{deluxetable*}{ccccccccc}
    \tablewidth{\textwidth}
    \tablecaption{Astrometric measurements from the literature used in the orbit fits presented in this paper. P.A.\ is the position angle. Here, RV$_{\rm pl}$ indicates a measurement of the planet's radial velocity \textit{relative to the primary}. Making this measurement involved separately measuring the absolute radial velocity of the planet (from a SINFONI spectrum) and the absolute radial velocity of the star (from a series of HARPS spectra), subtracting these quantities, and propagating the uncertainty.
    \label{tab:lit_astrom}}
    \tablehead{ Date & separation ($\rho$) & $\sigma_{\rho}$ & P.A. & $\sigma_{\rm P.A.}$ & RV$_{\rm pl}$ & $\sigma_{\rm RV_{\rm pl}}$ & instrument & reference\\ $[\mathrm{JD} - 2400000.5]$ & [mas] & [mas] & [deg] & [deg] & [\kms{}] & [\kms{}] & &}
\startdata
57538.4 & 830.4 & 4.9 & 150.28 & 0.22 & -- & -- & SPHERE & (1) \\
57565.5 & 830.1 & 3.2 & 150.14 & 0.17 & -- & -- & SPHERE & (1) \\
57791.0 & 827.6 & 1.5 & 150.11& 0.15 & -- & -- & SPHERE & (1) \\
57793.1 & 828.8 & 1.5 & 150.05 & 0.16 & -- & -- & SPHERE & (1) \\
57891.0 & 832 & 3 & 149.52 & 0.19 & -- & -- & NACO & (2) \\
57892.0 & 850 & 20 & 148.5 & 1.6 & -- & -- & NACO & (2) \\
58250.0 & 822.9 & 2.0 & 149.85 & 0.15 & -- & -- & SPHERE & (2) \\
58250.0 & 826.4 & 2.4 & 149.89 & 0.16 & -- & -- & SPHERE & (2) \\
58263.5 & -- & -- & -- & -- & 14 & 15 & SINFONI/HARPS & (3) \\
\enddata
\tablecomments{References: (1)~\citet{Chauvin:2017a},
(2)~\citet{Cheetham:2019a},
(3)~\citet{Petrus:2021a}.
}
\end{deluxetable*}

\subsection{Literature Data}

Literature spectral and photometric data used in this study (all of which are plotted in Figure \ref{fig:exorem_sed}) come from several sources. The VLT/SPHERE IFS spectra at Y-H ($0.96-1.64\,\mu$m) bands, four photometric VLT/SPHERE measurements at H2 ($\lambda_0=1.593\,\mu$m and $\Delta\lambda=0.11\,\mu$m) and H3 ($\lambda_0=1.667\,\mu$m and $\Delta\lambda=0.12\,\mu$m), at K1 ($\lambda_0=2.1025\,\mu$m and $\Delta\lambda=0.204\,\mu$m) and K2 ($\lambda_0=2.2550\,\mu$m and $\Delta\lambda = 0.218\,\mu$m) bands, and two NACO measurements at Lp ($\lambda_0 = 3.80\,\mu$m, $\Delta\lambda = 0.62\,\mu$m) and Mp ($\lambda_0=4.78\,\mu$m, $\Delta\lambda=0.59\,\mu$m) bands were originally published in \citet{Cheetham:2019a}. An additional NACO point obtained using the NB405 filter was originally published in \citet{Stolker:2020a}. The medium-resolution SINFONI spectrum at K band come from \citet{Petrus:2021a}. All spectral and photometric data (with the exception of the SINFONI data, which were kindly provided by S.~Petrus) were accessed from \texttt{species}\footnote{\url{https://github.com/tomasstolker/species/blob/main/species/data/companion_data.json}} (version 0.5.5). 

We also supplemented our new GRAVITY astrometry points with astrometric measurements from the literature (Table \ref{tab:lit_astrom}). These data were also taken verbatim from the papers indicated above. We used all of the available published astrometry in our orbit fits, with the exception of the JWST astrometry published in \citet{Carter:2022a}, and the NB405 NACO astrometry published in \citet{Stolker:2020a}. These points were excluded because they have large error bars relative to the other literature astrometric points (i.e. they contribute little additional information), and there is only a single astrometric point for each of these instrument/filter combinations (i.e. it is hard to tell whether they are affected by systematic errors). 

\section{Orbit Analysis}
\label{sec:orbit}

In order to understand how our new GRAVITY data constrains the orbit of HIP 65426 b, we performed seven different orbit-fits, varying the data subset used and the priors applied. These fits use literature data from Table \ref{tab:lit_astrom}\footnote{As mentioned in the previous section, there are additional astrometric datapoints given in \citet{Stolker:2020a} and \citet{Carter:2022a} which were excluded because of their large error bars.} and new VLTI/GRAVITY data from Table \ref{tab:astrom}. The results are summarized in Table \ref{tab:orbit_fits} and described in more detail below. We used version 2.2.2 of \texttt{orbitize!} (\citealt{Blunt:2020a}; \citealt{orbitize222}) for all fits, taking advantage of the ability to fit companion RVs introduced in version 2.0.0. \texttt{orbitize!} is a Bayesian tool for computing the posteriors over orbital elements of directly-imaged exoplanets, and is intended to meet the orbit-fitting needs of the high-contrast imaging community. All orbit fits used the following orbital basis: semimajor axis ($a$), eccentricity ($e$), inclination (inc, with $\mathrm{inc}=0$ for a face-on orbit), argument of periastron ($\omega_{\rm p}$), position angle of nodes ($\Omega$), and epoch of periastron, defined as fraction of the orbital period past a specified reference date (see \citet{Blunt:2020a} section 2.1 for a bit more detail). We use mjd=58859 as the reference date in this paper. Following the suggestion in \citet{Householder:2022a}, we also explicitly specify here that we assume that $\omega_{\rm p}$ is defined relative to the longitude of ascending node, which in turn is defined as the intersection point between the orbit track and the line of nodes when the object is moving \textit{away from the observer}. Our coordinate system is illustrated in Figure 1 of \cite{Blunt:2020a}. An interactive tutorial is also available to help users build an intuitive understanding of the coordinate system\footnote{\url{https://github.com/sblunt/orbitize/blob/main/docs/tutorials/show-me-the-orbit.ipynb}}. We used the same priors on all parameters as given in \citet{Blunt:2020a}, unless otherwise specified. 

\begin{figure*}
    \centering
    \includegraphics{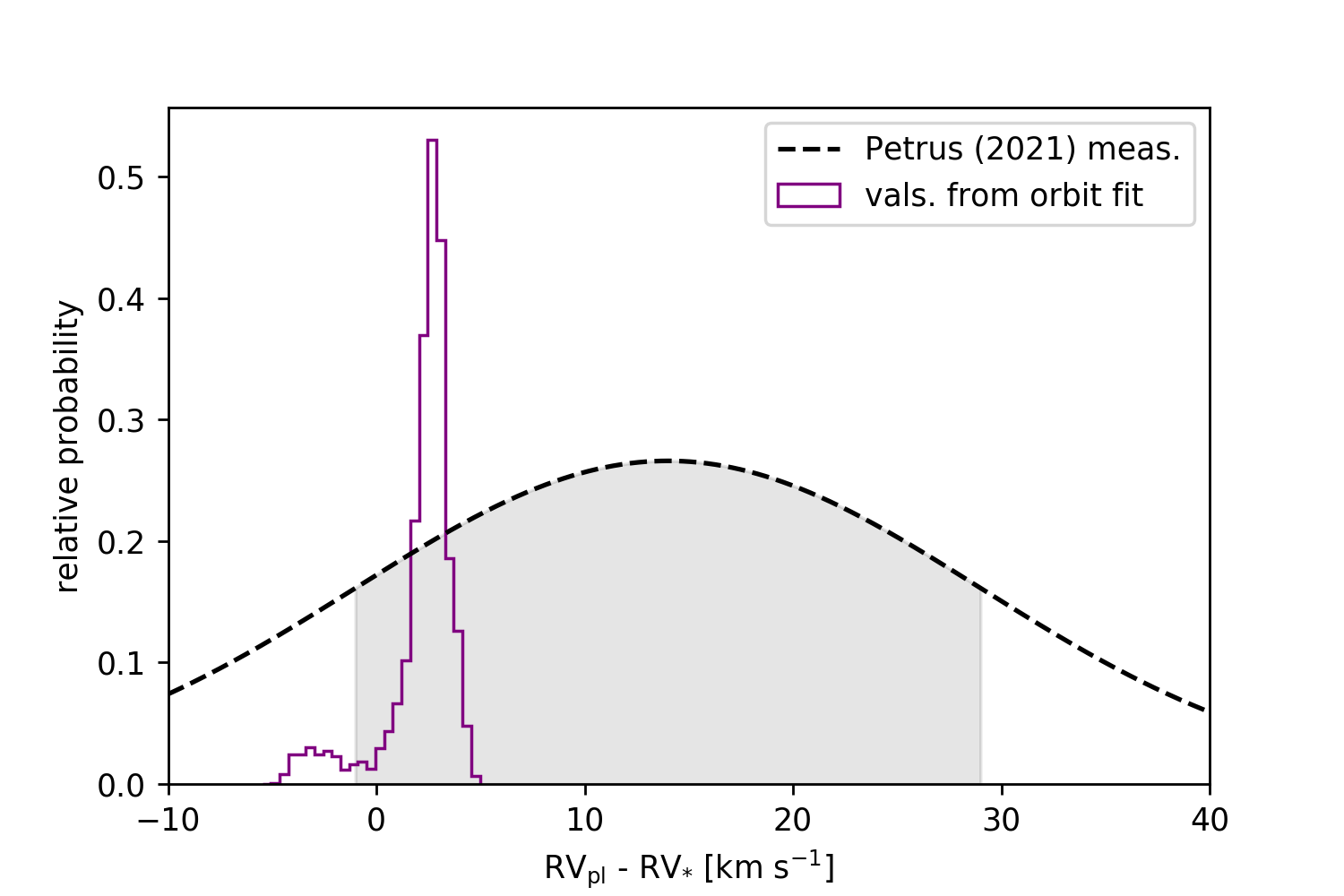}
    \caption{Planetary radial velocity predictions from the accepted fit (purple histogram), together with the planetary RV measurement from \citet{Petrus:2021a} (dashed black line, with 1-$\sigma$ range shaded grey). \textbf{Takeaway:} The planetary RV measurement does not constrain the orbital parameters, beyond breaking the 180$^{\circ}$ degeneracy for $\Omega$ and $\omega$. A planetary RV of $-3$~\kms{} is allowed, given the astrometry alone, but is disfavored (relative to a planetary RV of 3 \kms{}) because of the relative RV measurement.}
    \label{fig:rv}
\end{figure*}

The orbit fits presented in this paper include two types of input data: relative astrometry, and companion radial velocities. The ability to fit companion radial velocities is new in \texttt{orbitize!} since the publication of \citet{Blunt:2020a}, so we discuss it in a bit of detail here. Following standard practice (e.g. Chapter 1 of \citealt{Seager:2010a}), \texttt{orbitize!} by default defines orbital parameters (a, e, etc.) as those of the \textit{relative} orbit, meaning that astrometric and radial velocity measurements of the secondary are both assumed to be relative to the primary.\footnote{When radial velocity measurements are available for both the star and the planet, \texttt{orbitize!} instead assumes that planetary and stellar RV measurements are relative to the system barycenter.} In order to use the RV measurement of \citet{Petrus:2021a} in our orbit fits, we therefore subtracted the absolute measurement of the planet's RV (derived from the planetary spectrum) and an independent measurement of the star's RV derived from 78 HARPS spectra of the primary (see Section 4.2 of \citealt{Petrus:2021a} for details), and propagated the uncertainties in both measurements.\footnote{It is important to note that systematic offsets in the wavelength solutions used for the HARPS and SINFONI spectra (not to mention RV offsets due to astrophysical variability, such as pulsations) would affect the RV value derived here. However, the RV value derived from the SINFONI data is not sufficiently precise to impact our measurements of eccentricity and inclination, the main focus of this paper. Therefore we do not investigate any potential systematics in detail.} Relative companion RV measurements like this do not allow us to measure a dynamical mass for the companion, but have the potential to reduce posterior uncertainties of the relative orbit. However, the uncertainty on the available companion RV measurement is too large to meaningfully constrain the orbital parameters beyond the constraints from the relative astrometry (see Figure \ref{fig:rv}). It does, however, reduce the 180$^{\circ}$ degeneracy between $\Omega$ and $\omega$, which usually shows up as double-peaked posteriors on those parameters for relative-astrometry-only orbits. A more precise planetary RV would uniquely orient the planet in 3D space (to within the orbital parameter uncertainties). Figure \ref{fig:rv} shows the RV predictions for each orbit in the posterior of our accepted fit, together with the measurement from \cite{Petrus:2021a}. The current uncertainties of the companion and stellar RVs of HIP 65426 A and b limit the ability of the relative RV to constrain the orbit fit. 

We performed the following orbit fits (summarized in Table \ref{tab:orbit_fits}). All fits included the companion RV as described in the above paragraph.

\begin{enumerate}
    \item Only including literature data from SPHERE and NACO (i.e. no GRAVITY data)
    \item Literature data plus the first epoch of GRAVITY astrometry
    \item Literature data plus the second epoch of GRAVITY astrometry
    \item Literature data plus all GRAVITY astrometry
    \item The first two epochs of GRAVITY data alone (i.e. no literature data, and no third GRAVITY point)
    \item Fit 4, except fixing eccentricity to zero\footnote{Note that $\omega_p$ and $\tau_{58849}$ are undefined for a circular orbit.}.
    \item Fit 4, except applying a decreasing prior on eccentricity.
\end{enumerate}

For the final fit, we used the following prior distribution, following \citet{Nielsen:2008a}\footnote{The linear coefficient is from E.\ Nielsen (2023, private communication), and the constant value ensures the normalisation. Note that this prior has zero probability beyond $e=0.92$.
}:
\begin{equation}
\label{eq:ecc_prior}
    p(e) = - 2.18e + 2.01.
\end{equation}
Fits 1--5 were performed in order to assess the outlier sensitivity of our fits, as well as to understand the relative constraining power of the GRAVITY astrometry and the less precise literature measurements. The final two fits were performed to understand the prior dependence of the inferred eccentricity, as well as the impact of the eccentricity-inclination degeneracy (see, e.g., \citealt{Ferrer-Chavez:2021a}).

For all fits, we used the \texttt{ptemcee} implementation of \texttt{emcee} \citep{Vousden:2016a,Foreman-Mackey:2013a}, a parallel-tempered affine-invariant MCMC sampling algorithm. All runs used 20 temperatures and 1000 walkers. After an initial burn-in period of 100,000 steps, each walker was run for 200,000 steps. Every 100th step was saved.\footnote{In \texttt{orbitize!}, this configuration corresponds to the following variable definitions: \texttt{num\_temps = 20; num\_walkers = 1000; num\_steps = 200\_000\_000; burn\_steps = 100\_000; thin = 100}.} The chains were examined visually to determine whether they had converged, and the 200,000 steps of each chain post-burn-in were kept as the posterior estimate.

We applied Gaussian priors on parallax and total mass using the values given in Table \ref{tab:hip654}. Both parameters do not significantly correlate with any other fit parameters, and the marginalized posteriors on these parameters reproduce their priors (visible, for example, in Figure \ref{fig:corner}). 

As a final point, it is clear from scrutinizing Figure \ref{fig:orbit} that the two NACO points are discrepant from the contemporaneous SPHERE points. We briefly tested whether excluding the NACO points affected our fits by rerunning fit 6 without the two NACO points. Our results were indistinguishable, so we opted to keep the NACO points in all other fits.

\subsection{The Impact of GRAVITY}

\begin{figure*}
    \centering
    \includegraphics[width=\textwidth, trim={0, 0, 2cm, 0}]{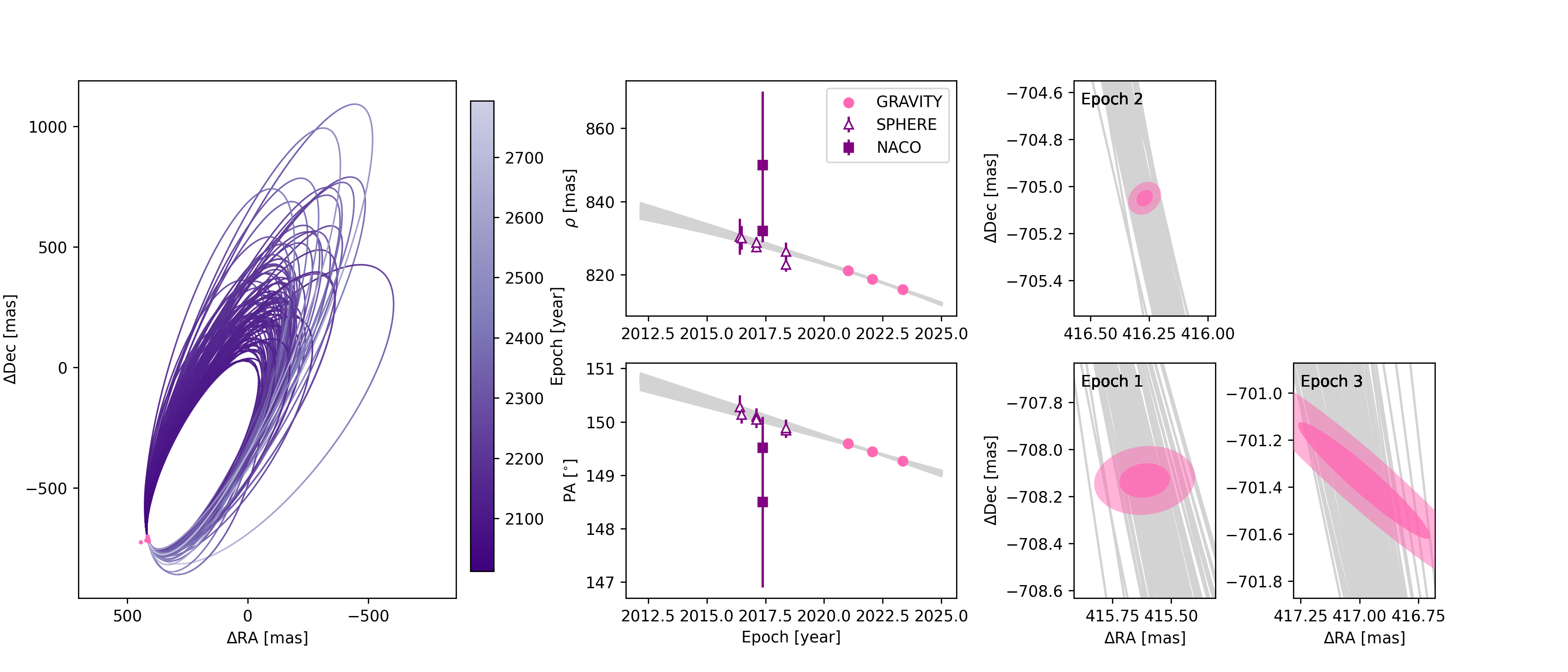}
    \caption{Sky-projected visualization of the posterior of the orbit fit \#4 described in the text. \textbf{Left:} 100 orbit tracks projected onto the plane of the sky, colored by elapsed time. The astrometric data are visible as pink points in the bottom left corner of the panel. \textbf{Middle column:} the same 100 posterior orbits (grey) in separation (top) and position angle (bottom) vs time, together with the astrometric data used for orbit-fitting. \textbf{Right column:} the same 100 posterior orbits (grey), together with earlier (bottom) and later (top) astrometric measurements taken with VLTI/GRAVITY. 1- and 2-$\sigma$ error ellipses are shaded in dark and light pink, respectively. \textbf{Takeaway:} The VLTI/GRAVITY epochs are $\sim$50x more precise than existing astrometric measurements of HIP 65426, and reduce the posterior uncertainty.}
    \label{fig:orbit}
\end{figure*}

\begin{figure*}
    \centering
    \includegraphics[width=\textwidth]{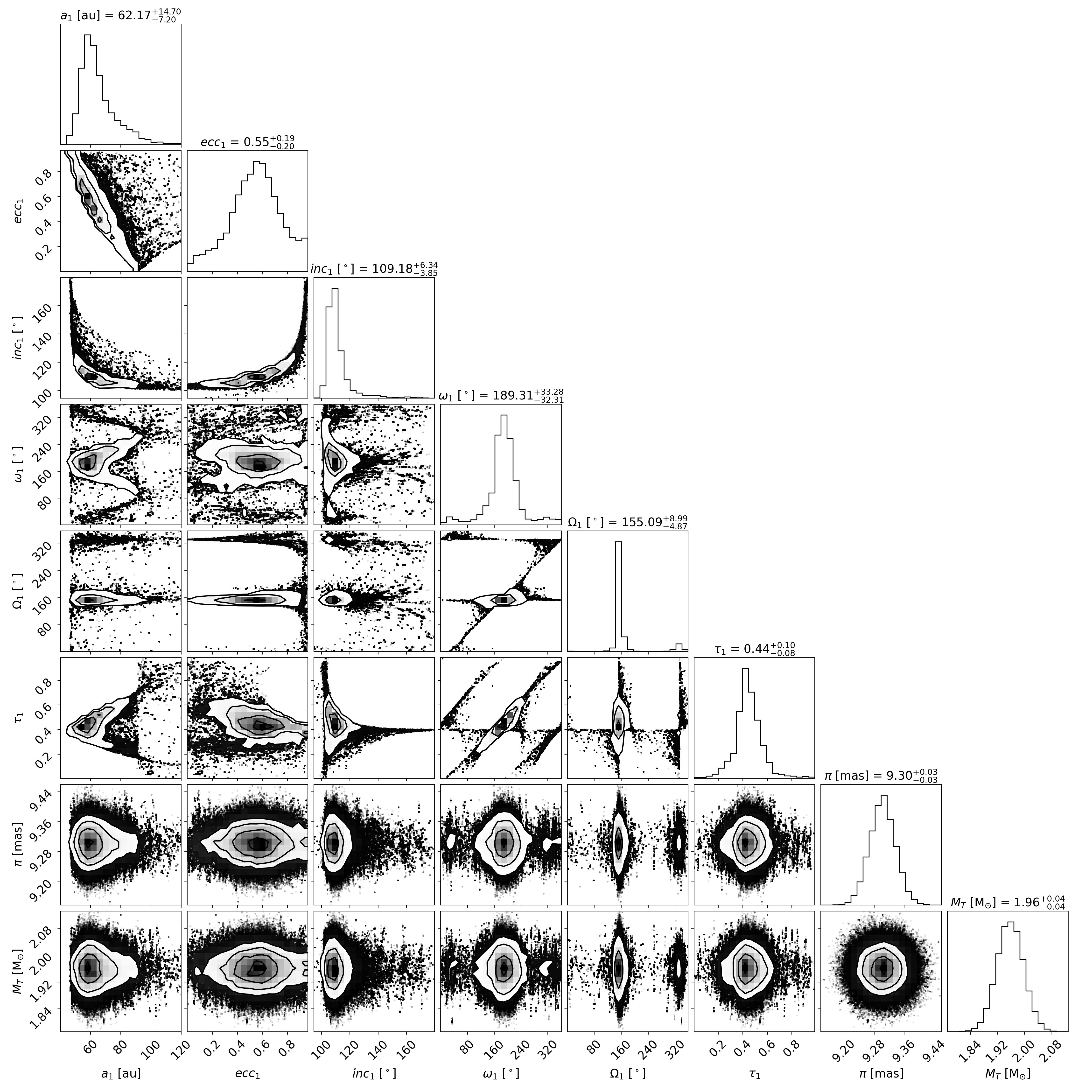}
    \caption{Corner plot of the posterior of the accepted orbit fit, using all data and assuming a uniform eccentricity prior. Diagonal panels show marginalized 1D histograms of posterior elements, and off-diagonals show 2D covariances between posterior elements. 1, 2, and 3-$\sigma$ contours are outlined in the covariance panels, and individual posterior samples outside of the 3-$\sigma$ boundaries are plotted directly as black dots. \textbf{Takeaway:} the 1D marginalized posterior distributions of semimajor axis and inclination are well constrained. Strong covariances are apparent, in particular between eccentricity, inclination, and semimajor axis.}
    \label{fig:corner}
\end{figure*}

\begin{figure*}
    \centering
    \includegraphics[width=\textwidth, trim={0, 2cm, 0, 5cm}]{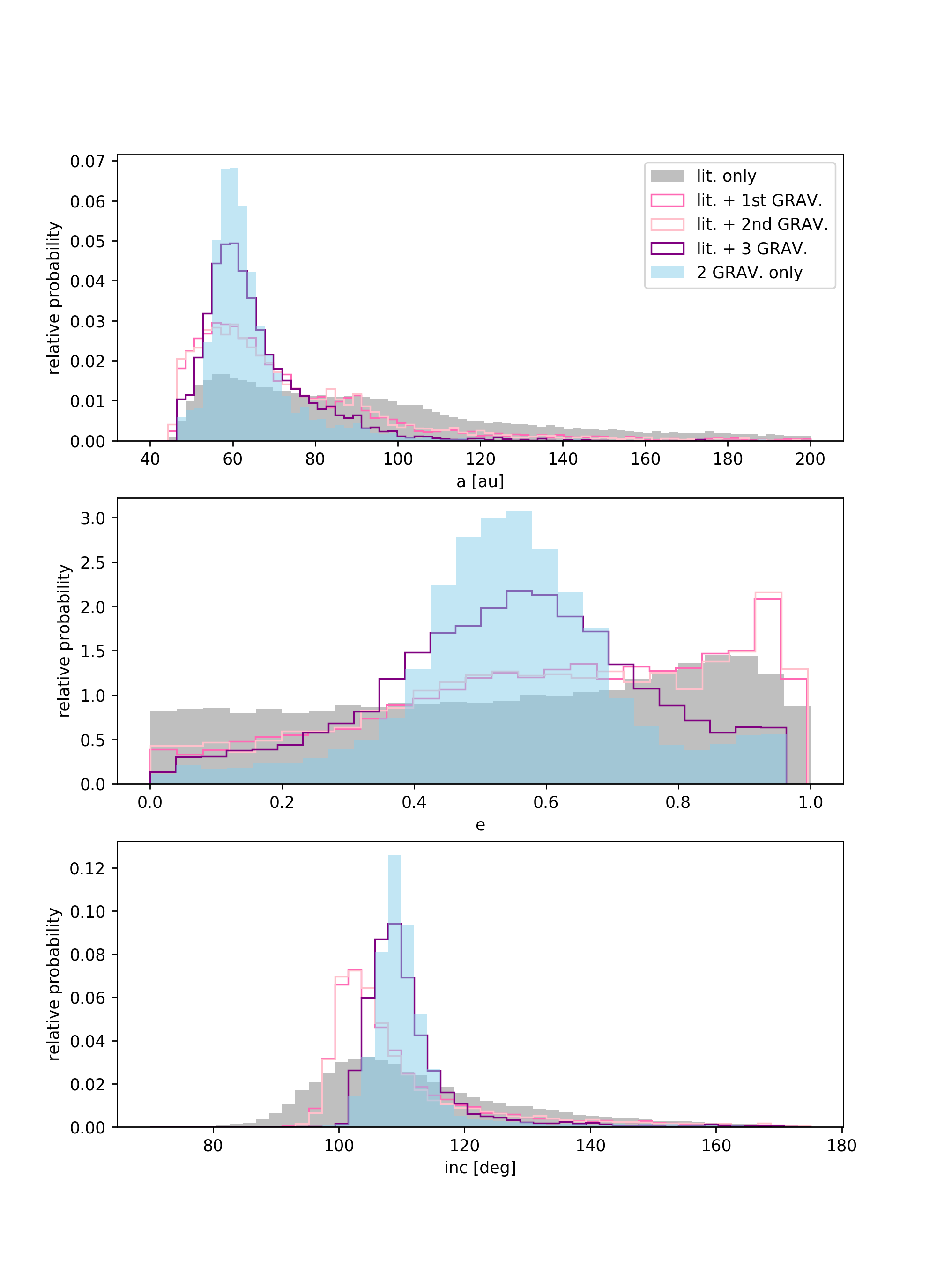}
    \caption{Relative constraining power of the astrometric data for semimajor axis (top), eccentricity (middle), and inclination (bottom). The results of the following fits are shown and compared: (1) only literature astrometry (i.e. no GRAVITY data; grey), (2) literature astrometry and the first epoch of GRAVITY data (dark pink outline), (3) literature astrometry and the second epoch of GRAVITY data (light pink outline), (4) only GRAVITY astrometry (i.e. no literature data), and (5) all astrometric data (i.e. the accepted fit; purple outline). \textbf{Takeaway:} most of the constraining power of the fit comes from the GRAVITY data, evident by the similarity between the GRAVITY-only fit and the accepted fit. In addition, neither GRAVITY point alone drives the fit, as evidenced by the similarity between fits (2) and (3). In other words, the posterior preference for moderate eccentricities is robust to the possibility that one of the two GRAVITY epochs is an outlier.}
    \label{fig:orbit_compare}
\end{figure*}

Figures \ref{fig:orbit} and \ref{fig:corner} visualize the posterior of fit \#4, using all astrometric data and assuming a uniform prior on eccentricity, and Figure \ref{fig:orbit_compare} compares each of the orbit fits that use varying data subsets. These plots show the transformative impact of GRAVITY data on the orbital uncertainties. The semimajor axis, eccentricity, and inclination posteriors all tighten significantly after including just three GRAVITY astrometric epochs. Figure \ref{fig:orbit_compare} also shows that the results of a preference for moderate eccentricity and near edge-on inclination are robust to outliers, by showing nearly identical posterior distributions regardless of which GRAVITY epoch is used in the fit. It is also apparent from Figure \ref{fig:orbit_compare} that the majority of the orbital parameter information is coming from the first two GRAVITY epochs, as evident by the similarity between the accepted fit and the GRAVITY-only fit. This test highlights the power of GRAVITY precision astrometry, but also warrants a warning: any unquantified systematics in the GRAVITY data could significantly change the recovered orbital parameters. The orbital constraints reported here rely heavily on the detailed work that the exoGRAVITY team has undertaken to extract accurate astrometry (e.g. \citealt{Lacour:2019b}, \citealt{GravityCollaboration:2020a}). HIP 65426 b should continue to be astrometrically monitored by GRAVITY in order to redistribute the impact of the first two (most precise) GRAVITY epochs reported here.

\subsection{A Moderate Eccentricity?}
\label{sec:ecc_interp}

\begin{figure*}
    \centering
    \includegraphics[width=\textwidth]{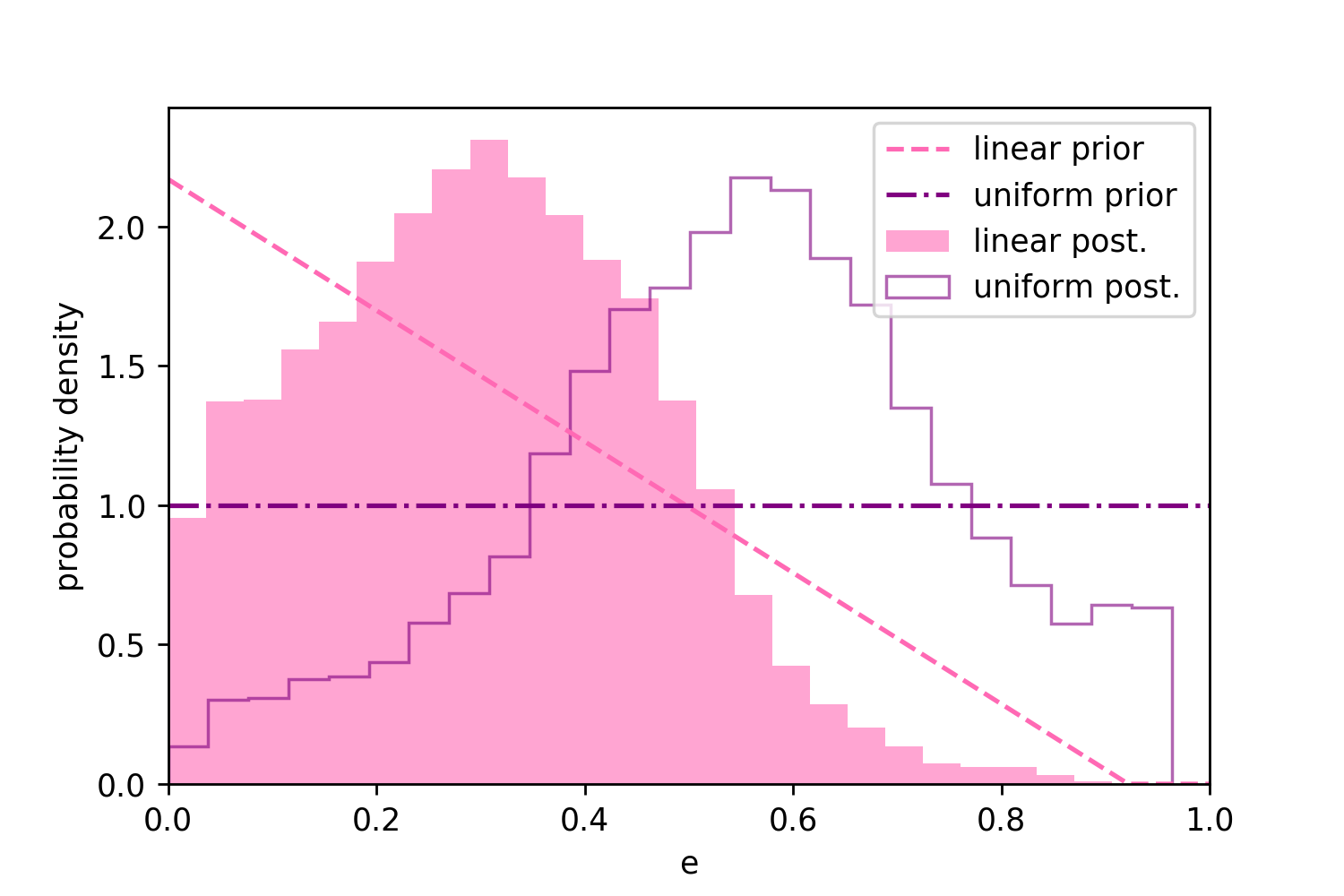}
    \caption{1D marginalized eccentricity posteriors for fits with uniform (purple) and linearly decreasing (pink) priors on eccentricity. The priors themselves are also plotted as lines of the same colors. \textbf{Takeaway:} the eccentricity posterior depends on the choice of prior. However, both the linearly decreasing prior and the uniform prior result in posterior peaks at moderate eccentricity values.}
    \label{fig:ecc_compare}
\end{figure*}

\begin{figure*}
    \centering
    \includegraphics{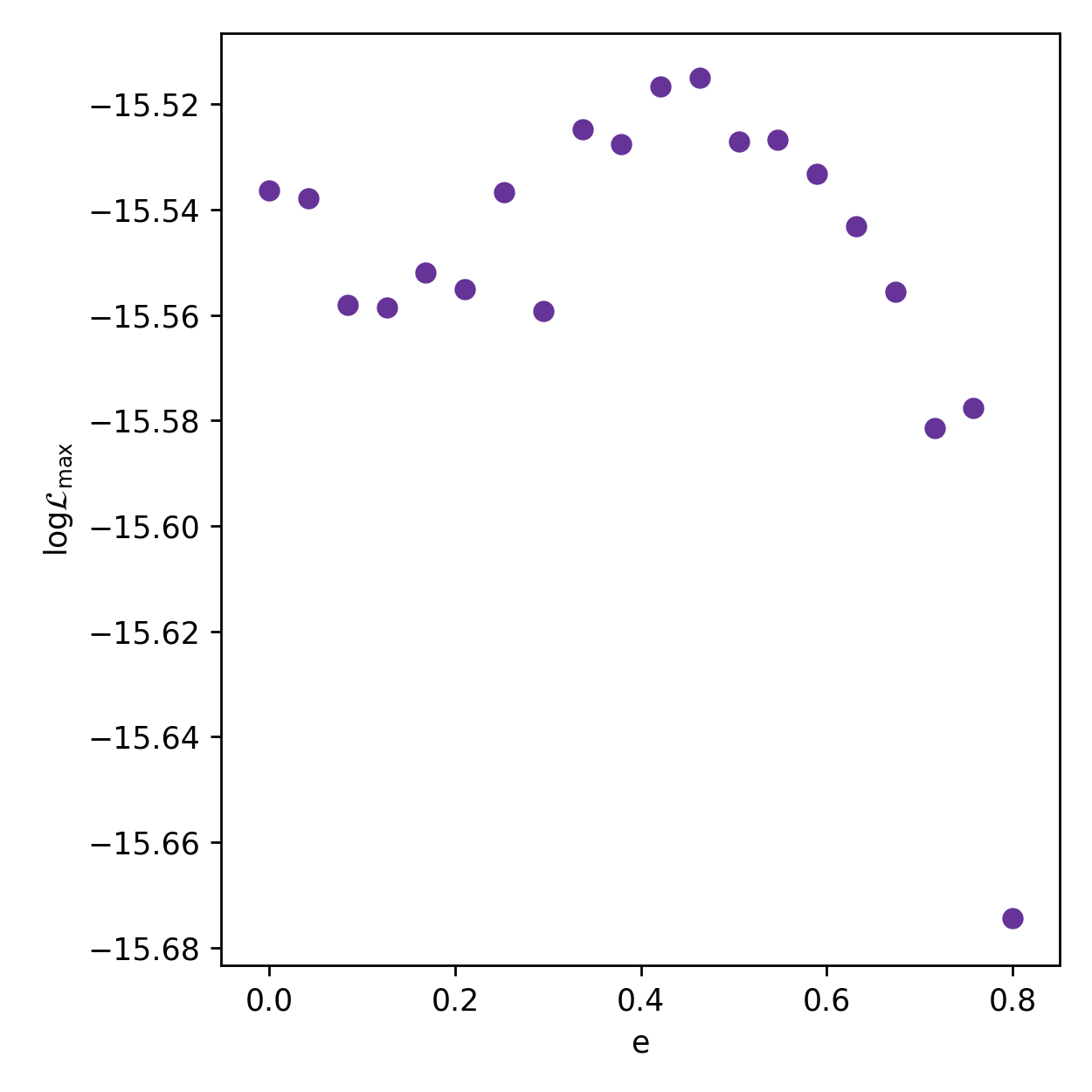}
    \caption{Maximum log(likelihood) as a function of eccentricity. Although the maximum a posteriori eccentricity is moderate ($\sim0.5$), the maximum likelihood at low and moderate eccentricities is comparable. This allows us to understand the shape of the eccentricity posterior (Figure \ref{fig:ecc_compare}); the likelihood is high at lower eccentricities, but the prior volume here is lower. The posterior ``drop-off'' at higher eccentricities is caused by a real decrease in likelihood. More eccentric orbits are less consistent with the data.}
    \label{fig:ecc_lnlike}
\end{figure*}

All of the orbit fits that include GRAVITY data show a preference for moderate eccentricities ($e\sim0.5$). In order to understand the significance of this preference, we performed a fit using a linearly decreasing eccentricity prior (Fit \#6; Equation \ref{eq:ecc_prior}). Figure \ref{fig:ecc_compare} shows that the eccentricity posterior is prior-dependent, even though both a linearly decreasing and a uniform eccentricity prior result in preferences for non-zero eccentricities. We next performed a series of maximum-likelihood fits, fixing eccentricity to a specific value for each, in order to examine how the changing maximum likelihood value was influencing the posterior shape. The results of this test are shown in Figure \ref{fig:ecc_lnlike}. This figure shows that the maximum likelihoods achieved by low (e $<$ 0.2) and moderate (e = 0.5) orbits are comparable, but that the highest achieveable maximum likelihood occurs at moderate eccentricities. This plot, along with Figures \ref{fig:corner} and \ref{fig:ecc_compare} allow us to construct the following explanation of the shape of the eccentricity posterior: at low ($e < 0.2$) eccentricities, likelihood is high, but prior volume is low (i.e. there are ``fewer'' circular orbits that fit the data well, even though those that do fit tend to fit very well). At moderate eccentricities, prior volume is high, and likelihood is also high, leading to a posterior peak. At high eccentricities ($e>0.6$), prior volume is very high, but likelihood is low, leading to a decreasing posterior probability as a function of eccentricity. The ``uptick'' at very high eccentricities ($e>0.8$) is caused by a degeneracy between eccentricity and inclination (see \citealt{Ferrer-Chavez:2021a} for a detailed discussion), together with a nonlinear relationship between eccentricity and sky projection. As the eccentricity asymptotically approaches 1, the corresponding inclination must increase more and more rapidly to reproduce the astrometric data (see the covariance between eccentricity and inclination in Figure \ref{fig:corner}). This causes a characteristic ``banana-shaped'' covariance, typical of incomplete orbits (see \citealt{Blunt:2019a} for another example in the context of an incomplete radial velocity orbit). We can explain the maximum a posteriori (MAP) shift to lower eccentricities when using a linearly decreasing prior, therefore, as occurring because the prior volume at moderate and high eccentricities decreases when we switch to a linearly decreasing eccentricity prior. This interpretation is supported by both the Akaike Information Criterion (AIC) and Watanabe-Akaike Information Criterion (WAIC) model selection metrics, a point which is explored further in the Appendix.

\begin{deluxetable*}{cccccccc}
    \tablewidth{\textwidth}
    \tablecaption{Marginalized posterior 68\% credible intervals for the free parameters in each of the orbit fits described in Section \ref{sec:orbit}. Parallax and total mass were additional free parameters in every. In all cases, the marginalized parallax 68\% credible interval was $9.30\pm{0.03}$ mas and the marginalized total mass 68\% credible interval was $1.96\pm{0.04}$ \msun{}. \label{tab:orbit_fits}}
    \tablehead{Fit name & a [au] & $e$ & inc [deg] & $\omega_{\rm p}$ [deg] & $\Omega$ [deg] & $\tau_{58849}$}
\startdata
Lit.\ data only & $92.5^{+88.7}_{-32.0}$ & $0.58^{+0.29}_{-0.38}$ & $109.5^{+20.5}_{-11.2}$ & $181.6^{+107.6}_{-128.9}$ & $171.9^{+163.9}_{-45.3}$ & $0.36^{+0.30}_{-0.17}$ \\
Lit.\ + GRAVITY epoch 1 & $66.1^{+28.3}_{-12.7}$ & $0.64^{+0.26}_{-0.31}$ & $105.6^{+13.9}_{-5.0}$ & $195.6^{+74.0}_{-95.9}$ & $156.9^{+171.0}_{-8.1}$ & $0.44\pm{0.15}$ \\
Lit.\ + GRAVITY epoch 2 & $66.3^{+27.2}_{-13.0}$ & $0.63\pm{0.27}$ & $105.4^{+14.9}_{-4.8}$ & $194.3^{+70.5}_{-101.6}$ & $156.5^{+172.2}_{-7.6}$ & $0.43\pm{0.15}$ \\
All data & $62.2^{+14.7}_{-7.2}$ & $0.55\pm{0.19}$ & $109.2^{+6.3}_{-3.9}$ & $189.3^{+33.3}_{-32.3}$ & $155.1^{+9.0}_{-4.9}$ & $0.44\pm{0.10}$ \\
GRAVITY epochs 1+2 only & $61.4^{+9.7}_{-5.1}$ & $0.55\pm{0.14}$ & $109.6^{+4.4}_{-2.9}$ & $190.3^{+25.8}_{-19.3}$ & $155.4^{+7.2}_{-3.6}$ & $0.44\pm{0.07}$\\
All data, fixed $e=0$ & $91.2\pm{0.4}$ & $\equiv0$ & $103.1\pm{0.5}$ & -- & $153.2^{+179.9}_{-0.3}$ & -- \\
All data, decreasing $e$ prior & $74.6^{+14.9}_{-10.0}$ & $0.30\pm{0.17}$ & $105.2^{+2.5}_{-1.9}$ & $187.6^{+50.5}_{-47.2}$ & $153.8^{+5.9}_{-2.5}$ & $0.46\pm{0.14}$ \\
\enddata
\end{deluxetable*}

\section{Spectral Analysis}
\label{sec:spectrum}

\begin{deluxetable*}{cccccccccc}
    \tablewidth{\textwidth}
    \tablecaption{68\% credible intervals of posterior fits to self-consistent model spectra grids. G is short for GRAVITY (i.e. the GRAVITY K-band spectrum was included in the fit), and Si is short for SINFONI (i.e. the SINFONI K-band spectrum was included in the fit). ``yes GP'' means that a Gaussian Process regression was enabled for the SPHERE IFS spectrum, the hyperparameter constraints for which are reported in the table. \label{tab:spectral_fits}}
    \tablehead{fit name & T$_{\rm eff}$ [K] & $\log{g}$ & $\pi$ [mas] & $l_{\rm SPHERE}$ [$\mu$m] & $A_{\rm SPHERE}$ & [Fe/H] & C/O & R [\rjup{}] & RV$_{\rm SINFONI}$}
\rotate
\startdata
\texttt{BT-Settl} (G only, yes GP) & $1637\pm8$ & $3.85\pm0.03$ & $9.3^{+0.03}_{-0.04}$ & $0.2^{+0.2}_{-0.1}$ & $0.4^{+0.3}_{-0.2}$ & --  & -- & $1.0\pm0.01$ & --\\
 & $1477^{+8}_{-7}$ & $3.93\pm0.07$ & $9.3\pm0.03$ & $0.3^{+0.1}_{-0.0}$ & $0.6\pm0.1$ & --  & -- & $1.19\pm0.03$ & --\\
\texttt{BT-Settl} (Si only, yes GP) & $1624^{+8}_{-7}$ & $3.89\pm0.04$ & $9.3\pm0.03$ & $0.4\pm0.1$ & $0.5\pm0.1$ & --  & -- & $0.99\pm0.01$ & $75395^{+1}_{-2}$\\
 & $1469^{+5}_{-4}$ & $3.71^{+0.07}_{-0.06}$ & $9.3\pm0.03$ & $0.4\pm0.1$ & $0.6\pm0.1$ & --  & -- & $1.24\pm0.02$ & $75396^{+1}_{-2}$\\
\texttt{BT-Settl} (G only, no GP) & $1639\pm7$ & $3.85\pm0.03$ & $9.3\pm0.03$ & -- & -- & --  & -- & $1.0\pm0.01$ & --\\
 & $1487^{+7}_{-6}$ & $3.96^{+0.05}_{-0.06}$ & $9.3\pm0.03$ & -- & -- & --  & -- & $1.17^{+0.03}_{-0.02}$ & --\\
\texttt{BT-Settl} (Si only, no GP) & $1630\pm7$ & $3.92\pm0.03$ & $9.31\pm0.03$ & -- & -- & --  & -- & $0.97\pm0.01$ & $75389^{+1}_{-2}$\\
 & $1475^{+5}_{-4}$ & $3.81^{+0.07}_{-0.06}$ & $9.3\pm0.03$ & -- & -- & --  & -- & $1.21^{+0.02}_{-0.03}$ & $75378\pm1$\\
\texttt{Exo-REM} (Si+G, yes GP) & $1337\pm9$ & $3.52^{+0.03}_{-0.02}$ & $9.3\pm0.03$ & $0.4\pm0.1$ & $0.5\pm0.1$ & $0.15^{+0.08}_{-0.1}$  & $0.595^{+0.008}_{-0.009}$ & $1.51^{+0.03}_{-0.02}$ & $18\pm8$\\
\enddata
\end{deluxetable*}

\newpage

The GRAVITY K-band spectrum (R=500) published in this work overlaps in wavelength coverage almost completely with the higher-resolution SINFONI spectrum (R$\sim$5600), and therefore does not add significant spectral information to the HIP 65426 b SED. However, it is an independent constraint on the K-band spectrum. In this section we first compare the GRAVITY and SINFONI spectra, finding good agreement once the GRAVITY spectrum has been rescaled as described in Section \ref{sec:data}, then globally compare the HIP 65426 b spectral data with self-consistent atmosphere models to update our understanding of this object's atmosphere. 

In Figure \ref{fig:gravity_and_sinfoni}, we plot the GRAVITY and SINFONI spectra. The agreement is excellent. Both spectra agree in magnitude (as expected after the rescaling) and shape. 

Following \citet{Petrus:2021a}, we derived atmospheric properties for HIP 65426 b by comparing with two sets of self-consistent atmospheric model grids: \texttt{BT-SETTL CIFIST 2011c}\footnote{\url{https://phoenix.ens-lyon.fr/Grids/BT-Settl/CIFIST2011c/}} (\citealt{Allard:2003a}, \citealt{Allard:2007a}, \citealt{Allard:2011a}) and \texttt{Exo-REM} (\citealt{Allard:2012a}, \citealt{Charnay:2018a}), excellent summaries of which are given in Section 3.1 of \citet{Petrus:2021a}. In the temperature range relevant for HIP 65426 b, the major advantages of \texttt{Exo-REM} include 1) ability to explore non-solar metallicities, and 2) the success of the \texttt{Exo-REM} cloud prescription at reproducing observations of dusty planets (like HIP 65426 b) near the L-T transition \citep{Charnay:2018a}. Ultimately, however, we are interested in comparing constraints from multiple independent models.

We used \texttt{species} \citep{Stolker:2020a} to perform comparisons to both sets of models. In all cases, we computed posteriors using the \texttt{pyMultinest} \citep{Buchner:2014a} Python interface to \texttt{multinest} (\citealt{Feroz:2008a}, \citealt{Feroz:2009a}, \citealt{Feroz:2019a}) with 1000 live points. The results of all atmosphere model fits are shown in Table \ref{tab:spectral_fits}. We also opted not to include rotational broadening as a free parameter in all cases, following an experiment showing that doing so resulted in a broad uniform marginalized posterior over the planet's vsini and unchanged other fit parameters. The spectral resolution of the SIFNONI spectra translate to a mininum detectable vsini of $\sim$50\kms{}, which translates to a rotation period of $\sim$0.1d, assuming i=90 and R=1.5\rjup{}, so the non-detection of rotational broadening is also consistent with our physical expectation. 

\subsection{\texttt{BT-SETTL CIFIST}}

\begin{figure*}
    \centering
    \includegraphics[width=\textwidth, trim = {2cm, 0, 3cm, 0}]{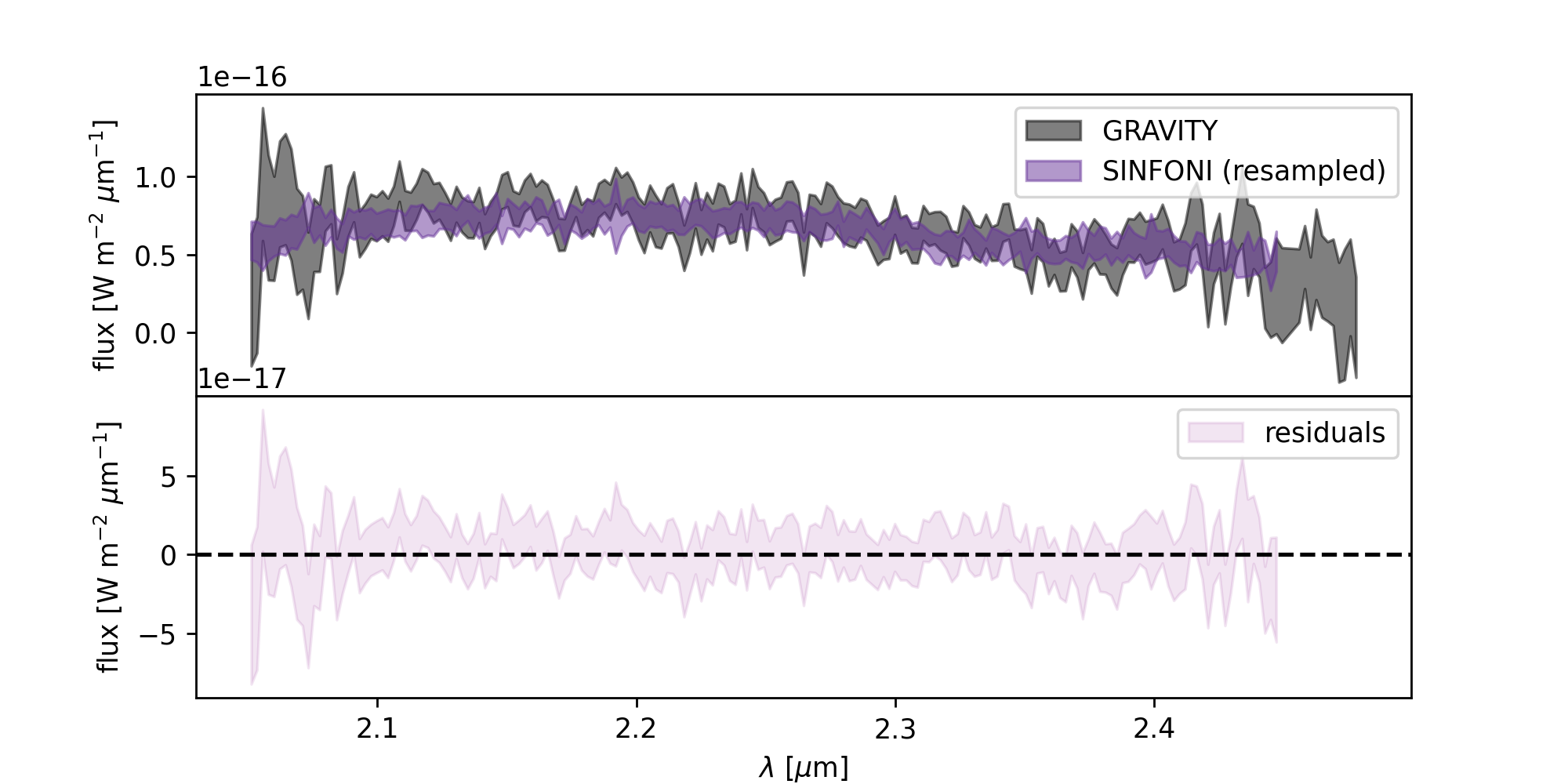}
    \caption{GRAVITY and SINFONI K-band spectra comparison. \textbf{Top:} GRAVITY (grey) and SINFONI (purple) 1$\sigma$ flux confidence intervals are shown as filled bands. The SINFONI spectrum was resampled onto the GRAVITY wavelength grid using \texttt{spectres} \citep{Carnall:2017a}. \textbf{Bottom:} The residuals, with propagated uncertainties, are shown relative to the flux=0 line. \textbf{Takeaway:} the agreement between these two independent datasets is excellent.}
    \label{fig:gravity_and_sinfoni}
\end{figure*}

\begin{figure*}
    \centering
    \includegraphics[width=\textwidth]{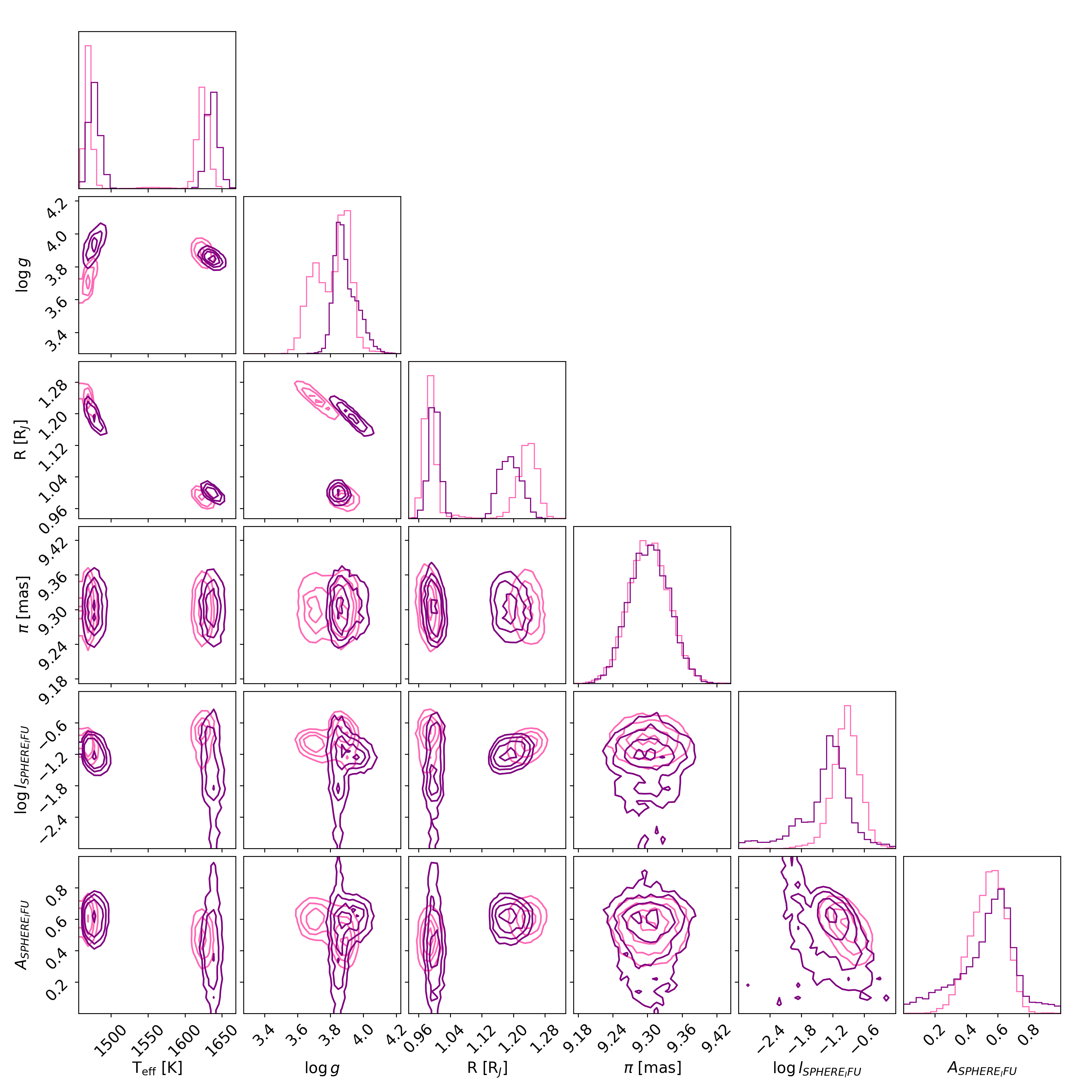}
    \caption{Results of forward-modeling the photometric and spectral data of HIP 65426 b by comparing with the \texttt{BT-SETTL CIFIST} model grid. Posteriors over the free parameters in the fit, as well radius, a derived parameter, are shown. Fits performed using GRAVITY K-band spectra are shown in purple, and fits performed using SINFONI K-band spectra are shown in pink. The GP hyperparameters (defined as in \cite{Wang:2020a} Equation 4) to the SPHERE IFS spectral data (length scale and amplitude) are shown as well. \textbf{Takeaways:} as expected, $\log{g}$ correlates strongly with radius and T$_{\rm eff}$. Two families of solutions are apparent at high (1.3 \rjup{}) and low (0.9 \rjup{}) radii.}
    \label{fig:btsettl_corner}
\end{figure*}

We performed four variations of \texttt{BT-SETTL CIFIST} comparisons to our full spectral dataset, by 1) using either the GRAVITY or SINFONI spectrum\footnote{Although we downsampled the resolution of the SINFONI spectrum in order to compare with the GRAVITY spectrum in Figure \ref{fig:gravity_and_sinfoni}, we fit the SINFONI spectrum at its native resolution.} in order to compare their relative constraining power, and 2) fitting for correlated noise in the SPHERE IFS data, following \citet{Wang:2020a} (see their Equation 4). All fits performed in this and the next section are summarized in Table \ref{tab:spectral_fits}. The results of the two fits allowing correlated noise in the SPHERE data are shown in Figure \ref{fig:btsettl_corner}. The fit only including GRAVITY data is shown in purple, and the fit only including SINFONI data is shown in pink. The two fits are consistent overall, as we would expect given the consistency of the spectra themselves. Like \citet{Petrus:2021a} (but unlike \citealt{Carter:2022a}), we also recover a bimodal posterior in surface gravity and effective temperature, regardless of which of the two K-band spectra we use. 

\begin{figure*}
    \centering
    \includegraphics[width=\textwidth]{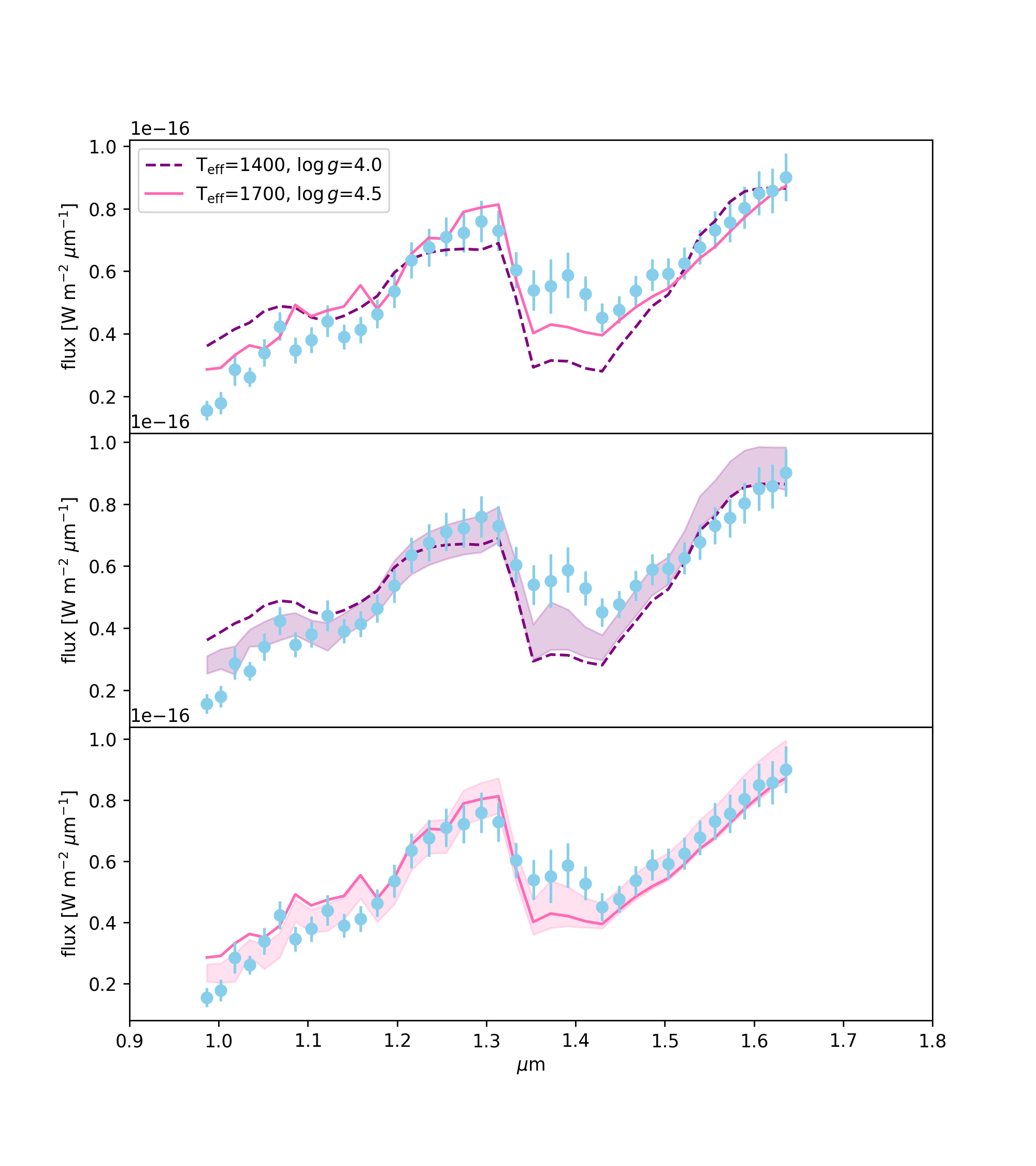}
    \caption{\texttt{BT-SETTL CIFIST} models representing the two posterior peaks shown in Figure \ref{fig:btsettl_corner}, together with the SPHERE IFS data. \textbf{Top:} both models, resampled onto the SPHERE IFS wavelength grid using \texttt{spectres} \citep{Carnall:2017a}, and multiplied by a scalar chosen to minimize the sum of squared residuals for the SPHERE IFS data alone. The SPHERE IFS data are shown as blue points, with error bars representing their reported statistical uncertainties. \textbf{Middle:} the low-T$_{\rm{eff}}$ model (dashed purple line), SPHERE IFS data, and Gaussian Process 1$-\sigma$ uncertainties (solid purple band; computed using the MAP GP parameters). \textbf{Bottom:} same as middle, but the high-T$_{\rm{eff}}$ model is shown as a solid pink line, and GP uncertainties as a pink band. \textbf{Takeaway:} there are correlated residuals in the SPHERE passband for both of the T$_{\mathrm{eff}}$ modes recovered from comparisons to the \texttt{BT-Settl CIFIST} grid, which we model with a squared-exponential Gaussian process.}
    \label{fig:sphere}
\end{figure*}

Our initial \texttt{BT-SETTL CIFIST} fits allowed a free RV offset parameter for the SINFONI data, but this value consistently converged to unphysically large values, perhaps indicating an underlying problem with the model grid. We therefore opted to fix the RV of the SINFONI data to 0 for the fits presented here. Toggling on and off a GP for the SPHERE data does not significantly impact the physical parameters inferred, even though there are clear correlated residuals in the SPHERE data (Figure \ref{fig:sphere}). The SPHERE residuals are visualized in Figure \ref{fig:sphere}, where both modes of the posterior are plotted along with the SPHERE IFS data. Toggling on a GP for this dataset allows the model to treat the deviations from both models as correlated noise, perhaps due to imperfect speckle subtraction at the planet location, photometric calibration errors due to telluric absorption, and/or model imperfections.

\subsection{\texttt{Exo-REM}}

Because the agreement between the \texttt{BT-SETTL CIFIST} model fits is good regardless of whether we use the SINFONI or GRAVITY spectrum, we only perform \texttt{Exo-REM} fits using \textit{both} datasets. 

\begin{figure*}
    \centering
    \includegraphics[width=\textwidth]{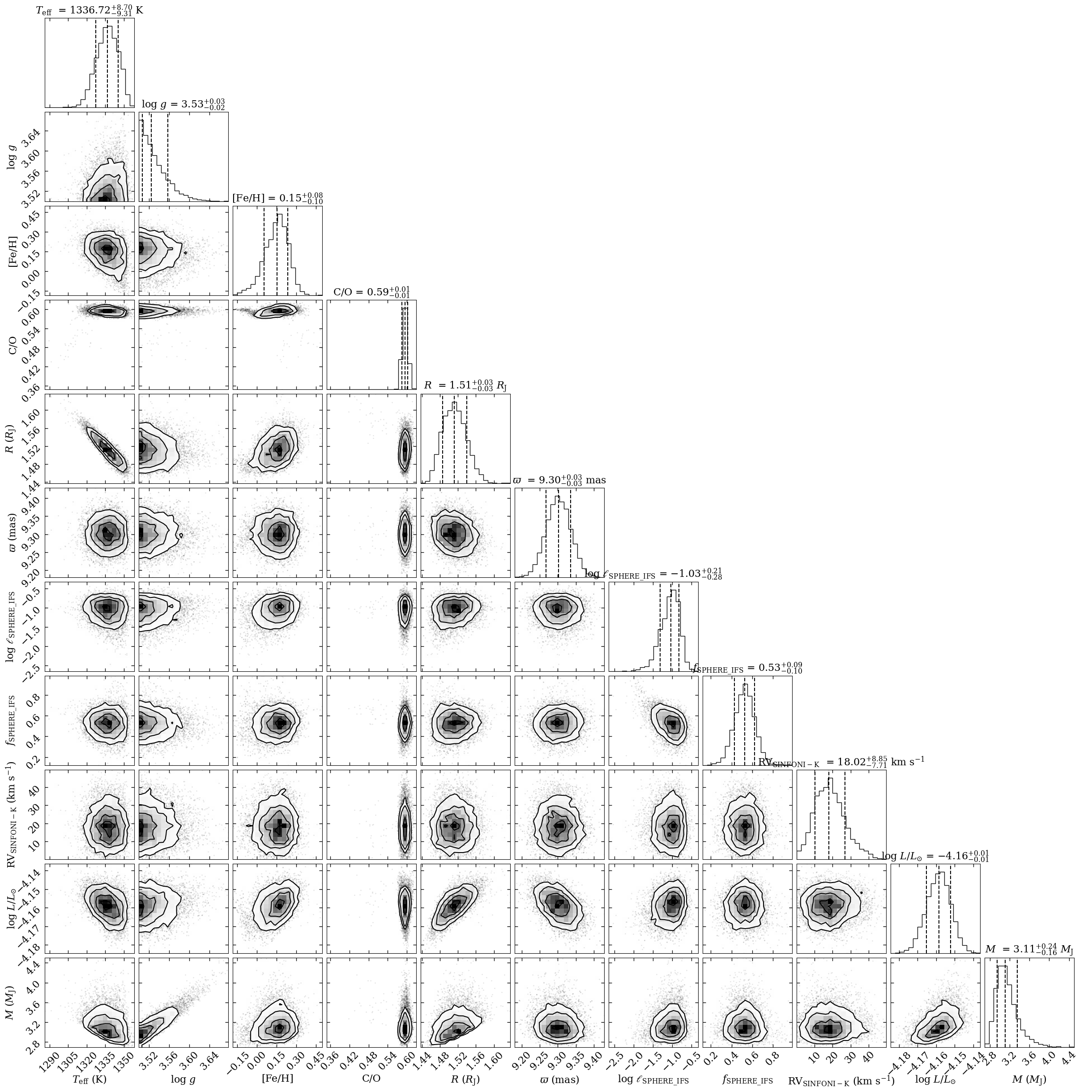}
    \caption{\texttt{Exo-REM} posterior fits to all data, showing 2D covariances and 1D marginalized posteriors over fitted model parameters and derived parameters (radius, luminosity, and mass). Also see Figure \ref{fig:exorem_sed}. Note: parallax is denoted $\varpi$ here, and $\pi$ elsewhere in the text. \texttt{Takeaways:} the \texttt{Exo-REM} derived atmosphere parameters are about 150K lower than the low-T$_{\rm eff}$ \texttt{BT-SETTL CIFIST} parameters (Figure \ref{fig:btsettl_corner}). The surface gravity hits the edge of the available grid. A slightly super-solar-metallicity atmosphere with a C/O ratio of 0.6 is favored, although there are likely systematic errors unaccounted for in this fit.}
    \label{fig:exorem_corner}
\end{figure*}

\begin{figure*}
    \centering
    \includegraphics[angle=90, height=8.5in]{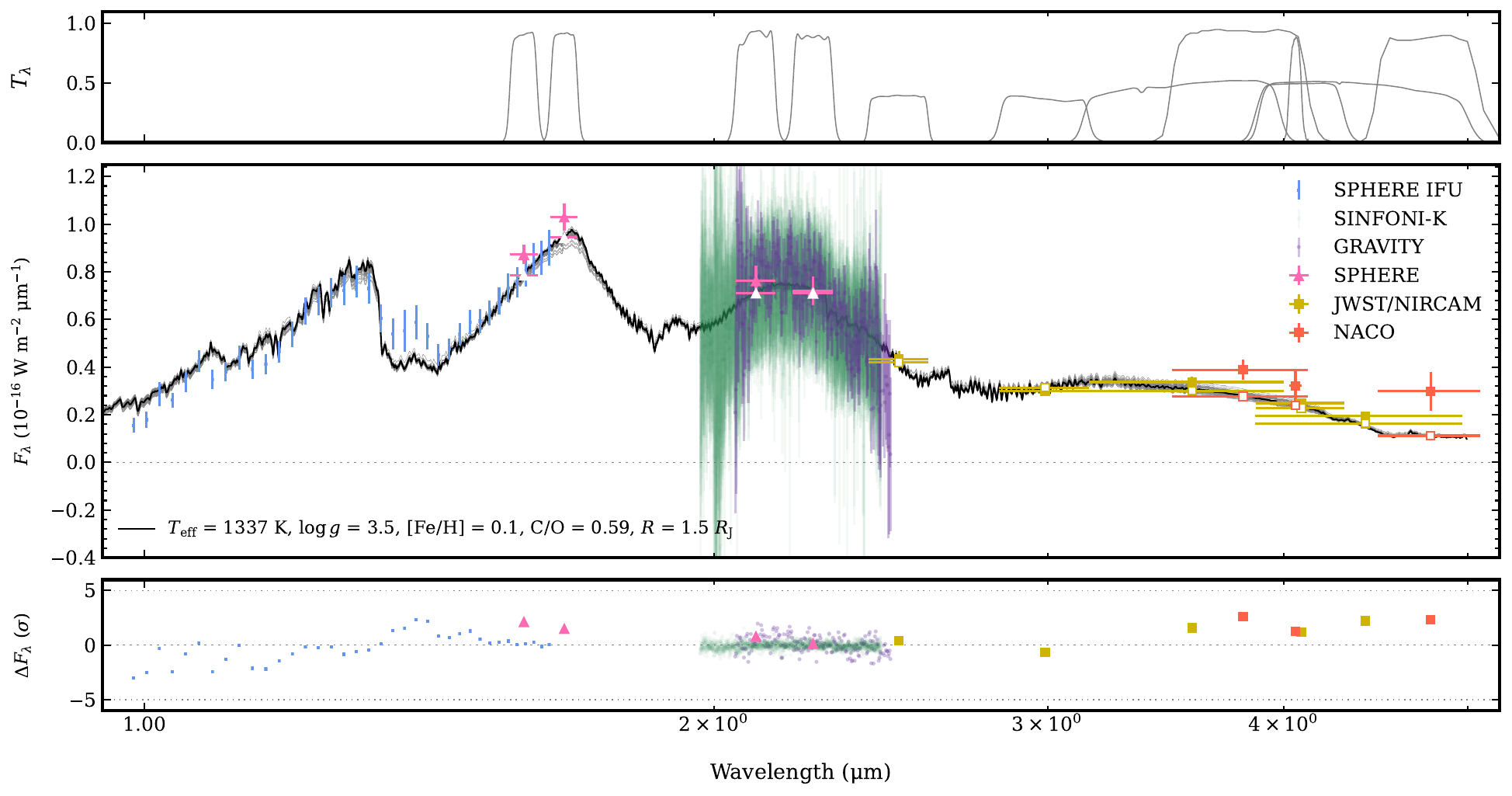}
    \caption{Full best-fit model SED from comparison with the Exo-REM grid, together with all fitted spectral data. \textbf{Top}: transmission functions of photometric bands. \textbf{Middle:} MAP model spectrum, along with 100 random draws from the posterior (in grey; barely distinguishable from MAP spectrum). The spectral data are plotted as points with error bars. The horizontal bars of the photometry points indicate their spectral bandpass. The corresponding band-integrated model predictions are overplotted as empty symbols. \textbf{Bottom:} MAP model residuals. \textbf{Takeaways:} Overall, the model spectrum fits well. Correlated noise is visible in the residuals of the SPHERE dataset. The NACO and JWST/NIRCAM points beyond 3 $\mu$m are underestimated by the model.}
    \label{fig:exorem_sed}
\end{figure*}

The available \texttt{Exo-REM} grid is computed for a grid of metallicities, allowing us estimate the atmospheric metallicity and C/O ratio. However, the grid we used only computes predictions out to 5 $\mu$m, so we excluded the two longer-wavelength JWST/MIRI photometry for these fits. The results are shown in Figures \ref{fig:exorem_corner} and \ref{fig:exorem_sed}, and summarized in Table \ref{tab:spectral_fits}. A solar metallicity with a C/O ratio of 0.6 is preferred. We find GP parameters for the SPHERE dataset that are consistent with the results from the \texttt{BT-Settl CIFIST} fits (previous section), a good sanity check that the model is similarly treating correlated noise in both cases. We also recover a planetary radial velocity value consistent with that reported in \cite{Petrus:2021a}, which is an excellent sanity check given that their planetary RV was computed by cross-correlation with the SINFONI spectrum alone, and not through SED fitting. Unsurprisingly, our lower-resolution GRAVITY spectrum does not permit a direct RV measurement. The effective temperature recovered by this fit is about 150K lower than the low-temperature \texttt{BT-SETTL CIFIST} mode, increasing our overall confidence in this low-temperature interpretation. However, the surface gravity value pushes against the low-gravity end of the grid, which casts doubt on the results of this fit.

The constraints we derived in this analysis are significantly more precise than those reported in \citet{Petrus:2021a}, because at the time of that paper's publication the \texttt{Exo-REM} grid predictions were only available for K-band. We were able to use the updated \texttt{Exo-REM} grid to compare with all available spectral information. However, the uncertainties are likely underestimated, in particular because these fits do not account for interpolation errors, an important point that we discuss in more detail in the next section.

From scrutinizing the residuals shown in Figure \ref{fig:exorem_sed}, it is also clear that the NACO and JWST photometry beyond 3 $\mu$m is systematically higher than the \texttt{Exo-REM} models. Therefore, in addition to potentially underestimated errors due to un-modeled physics in the atmosphere grid, interpolation errors, and correlated noise in the SPHERE IFS spectrum, the best fit is still not perfect, again pointing to potential inaccuracies in our comparison. Future work could attempt to fit offsets between photometric data from different instruments in order to reduce the discrepancy, and/or inflate the errors in these or other spectral datapoints. Altogether, the values and errors derived from grid comparison should be treated with caution.

\section{Discussion \& Conclusion}
\label{sec:discuss}

\subsection{Interpreting the Eccentricity Constraints}

Because of the strong degeneracy between eccentricity and inclination, it is most straightforward to report constraints on these parameters in two dimensions. The direction of orbital motion on the plane of the sky constrains the inclination to $>90^{\circ}$. For fit \# 4, including all available astrometry and applying a uniform eccentricity prior, the 1-$\sigma$ upper limit on both parameters is $e=0.7$/inc=110$^{\circ}$, and the 2-$\sigma$ upper limit is $e=0.8$/inc=120$^{\circ}$. Applying a non-uniform, linearly decreasing prior on eccentricity, which previous work shows is appropriate for the cold Jupiter population (\citealt{Bowler:2020a}, \citealt{Nagpal:2023a}), tightens these upper limits even more. These are still tenuous constraints, but they are driven by the likelihood, not just by the prior. This is the first time we are obtaining eccentricity posteriors on this object that do not simply reproduce the prior. It is worth emphasizing the difficulty of measuring the eccentricity of an object almost 100 au from its star; previous studies of this object report that it would take 5-10 years of orbit monitoring before resolving orbital curvature and constraining HIP 65426 b's eccentricity. With the precision of VLTI/GRAVITY, we are able to ``speed up time'' and obtain eccentricity constraints sooner, with fewer measurements. Other directly imaged exoplanets with well-constrained eccentricities are generally much closer to their stars (e.g. $\beta$ Pic b and c, at 3 and 10 au).

A potentially useful outcome of this paper is a generalizable prescription for interpreting eccentricity posterior for incomplete orbits. To have confidence in an eccentricity measurement from a posterior, we suggest showing that the eccentricity posterior is:
\begin{itemize}
    \item prior-independent, that is, driven by the likelihood and not phenomenologically different under different prior assumptions,
    \item inconsistent with a circular orbit, ideally beyond 3-$\sigma$ for varied prior assumptions, and
    \item not dependent on a single data point.
\end{itemize}

HIP 65426 b does not yet satisfy the first two criteria, so continued orbital monitoring with VLTI/GRAVITY will be important to robustly measure its eccentricity. Planetary RVs with sub-\kms{} precision (Figure \ref{fig:rv}), which could be obtained with high spectral resolution instruments like CRIRES, would also further constrain the eccentricity (e.g., \citealt{Snellen:2014a}, \citealt{Schwarz:2016a}).

With these caveats in mind, the posteriors presented in this paper favor a low or moderate eccentricity. Given the results of \cite{Marleau:2019a}, this is a preliminary hint that HIP 65426 b did not attain its current position via scattering after disk dispersal.

\subsection{Interpreting the Atmosphere Constraints}

In this analysis, we have focused on updating the atmospheric model analysis, independent from the evolution-based analysis of \citet{Carter:2022a}, which constrained physical atmospheric properties using an estimate of L$_{\mathrm{bol}}$. First, it is important to understand the limits of the spectral interpretation approach we have taken in this paper. Self-consistent grid modeling involves interpolating spectra in multiple dimensions during the spectral inversion, while the variations of the synthetic spectra along the grid are not linear in these dimensions (e.g. \citealt{Czekala:2015a}, \citealt{Petrus:2021a}). Missing or incorrect physics in the model grids, therefore, is not the only source of error. Performing atmospheric retrievals to evaluate the grid comparison results of this study is an important next step.

Keeping this limitation in mind, in this work we repeated the analysis presented in \citet{Petrus:2021a}, which compared the available spectral and photometric data of HIP 65426 b with the self-consistent \texttt{BT-Settl} and \texttt{Exo-REM} grids. This work benefited from additional data (in particular, the medium-resolution K-band GRAVITY spectrum, an additional NACO photometric point presented in \citealt{Stolker:2020a}, and the JWST photometery presented in \citealt{Carter:2022a}), and the expanded capability of \texttt{Exo-REM} to handle data outside of K-band. Like \citet{Petrus:2021a}, we recovered two modes in our \texttt{BT-Settl} posterior fit, regardless of which K-band spectrum we use and whether we included a correlated noise model for the SPHERE IFS data: one at a higher radius of 1.2 \rjup{}, and one at a lower radius of 1.0 \rjup{}. Both modes are significantly below the hot-start radius of 1.4 \rjup{} derived in \cite{Carter:2022a}, reinforcing the tension that paper originally pointed out.

Like \citet{Petrus:2021a}, we only recover a single posterior mode when comparing with the \texttt{Exo-REM} grid, which is about 150K cooler than the coolest MAP \texttt{BT-Settl} fit. The Gaussian process hyperparameters applicable to the SPHERE datasets are consistent across both grids, which we interpret as the presence of correlated observational noise, but which could also be a systematic problem common to both grids. The \texttt{Exo-REM} fit posterior favors a slightly super-solar metallicity and a C/O ratio of 0.6. However, the $\log{g}$ posterior hits the edge of the grid, so we strongly encourage skepticism of these derived values. The C/O ratio, metallicity, and other atmosphere properties we present here are consistent but more precise than those presented in \citet{Petrus:2021a}, keeping in mind that the systematic errors are likely underestimated.

A planetary metallicity \textit{relative} to its host star's is generally expected to reflect its formation condition (\citealt{Oberg:2011a}, \citealt{Madhusudhan:2014a}); in particular, formation via gravitational instability is generally held to produce an object with the same metallicity as its primary. In order to interpret HIP 65426 b's metallicity, it is important to understand the primary's metallicity. Because HIP 65426 A is a fast rotator with few spectral lines, a direct metallicity measurement is difficult, but if we take the metallicity of other members of its moving group as representative, it should be approximately solar, modulo scatter among individual stars (see Section 4.3 of \citealp{Petrus:2021a}).

Taken with a big grain of salt, the super-solar metallicity of HIP 65426 b, expected $\sim$solar metallicity of HIP 65426 A, and the planetary C/O ratio of 0.6, as constrained by comparison with the \texttt{Exo-REM} grid, are consistent with formation via core accretion past the CO snowline (\citealt{Oberg:2011a}). The exact location of this snowline for HD 163296, which has a similar mass to HIP 65426 A, was observed to be 75 au (\citealt{Qi:2015a}, \citealt{Petrus:2021a}). This (very tentative, with many caveats, see, e.g., \citealt{Molliere:2022a}) picture provides a parallel constraint on the picture of how HIP 65426 b attained its current separation by setting an outer limit on the initial formation location before any scattering occurred. 

\subsection{Future Directions}

Continued orbital monitoring, both with VLTI/GRAVITY and spectrographs capable of measuring additional planetary RVs will refine the eccentricity measurement of HIP 65426 b over the next few years, allowing us more insight into this specific planet's formation and further constraining the population-level eccentricity distribution of cold Jupiters. Uncertainties of $\sim\kms{}$ or below are needed for relative RV measurements of HIP 65426 b to constrain orbital parameters like a and e, motivating observation with high resolution spectrographs like CRIRES. Other tracers of formation condition that will be measurable in the near future include the planet's obliquity \citep{Sepulveda:2023a}, dynamical mass (hopefully measurable upon the release of Gaia timeseries data), and spin (from high resolution spectra). Further atmospheric characterization work, particularly retrievals that jointly model bulk atmosphere properties and trace chemical species fingerprints (see, e.g. \citealt{Xuan:2022a}) represent an important parallel path toward assessing the trustworthiness of the metallicity and C/O ratio, which will be helpful for pinpointing the formation location of HIP 65426 b. \citet{Xuan:2022a} showed that high resolution spectra are sensitive to a broader range of atmospheric pressures than lower-resolution spectra, leading to more robust abundance measurements, motivating high resolution measurements of HIP 65426 b in particular.

It is worth mentioning that the rapid rotation and early spectral type\footnote{Hot, early-type stars have few spectral lines, and rapid rotators have broad spectral lines, which both decrease RV precision.} of the primary star, HIP 65426 A, precludes precise RV measurements, so a dynamical mass measurement will rely entirely on the combination of relative (i.e. from high contrast imaging) and absolute (i.e. from Gaia) astrometry. However, radial velocity monitoring of the planet, together with continued relative astrometric monitoring, may allow for the detection of unseen inner companions, through Keplerian effects on the host star \citep{Lacour:2021a} and/or planet-planet interactions \citep{Covarrubias:2022a}. 

More theoretical work is also needed in order to interpret these results. In particular, population-level studies \`a la \cite{Marleau:2019a} could be conducted for alternate plausible formation pathways, particularly more rapid core formation via pebble accretion and gravitational instability in the protoplanetary disk. It would also be interesting to compare the existing measurements of other formation tracers, for example orbital inclination, metallicity, and C/O ratio, with their corresponding predictions in existing models.

VLTI/GRAVITY is a powerful instrument. This study has mostly focused on its ability to refine orbital eccentricity measurements in order to make dynamical inferences useful for commenting on planet formation. However, the ExoGRAVITY program is actively investigating a number of other scientific questions and observational constraints, particularly precise dynamical mass measurements and bolometric luminiosities (e.g., \citealt{Hinkley:2023a}).

Planet formation is a complex process, and we will need a diverse set of observational and theoretical tools to unravel its secrets. This paper represents a small step toward better constraints and deeper understanding.

\begin{acknowledgements}

We thank Aldo Sepulveda and Dan Huber for collaborative and informative discussions.
S.B.\ and
J.J.W.\ are supported by NASA Grant 80NSSC23K0280.
G.-D.M.\ acknowledge the support of the DFG priority program SPP 1992 ``Exploring the Diversity of Extrasolar Planets'' (MA~9185/1) and from the Swiss National Science Foundation under grant
200021\_204847 ``PlanetsInTime.''
Parts of this work have been carried out within the framework of the NCCR PlanetS supported by the Swiss National Science Foundation.
S.-P. acknowledges the support of ANID, – Millennium Science Initiative Program – NCN19\_171.
S. L. acknowledges the support of the French Agence Nationale de la Recherche (ANR), under grant ANR-21-CE31-0017 (project ExoVLTI).
This work is based on observations collected at the European Southern Observatory under ESO programme 1104.C-0651. It also made use of data from the European Space Agency (ESA) mission
{\it Gaia} (\url{https://www.cosmos.esa.int/gaia}), processed by the {\it Gaia}
Data Processing and Analysis Consortium (DPAC,
\url{https://www.cosmos.esa.int/web/gaia/dpac/consortium}). Funding for the DPAC
has been provided by national institutions, in particular the institutions
participating in the {\it Gaia} Multilateral Agreement.
This publication makes use of data products from the Wide-field Infrared Survey Explorer, which is a joint project of the University of California, Los Angeles, and the Jet Propulsion Laboratory/California Institute of Technology, funded by the National Aeronautics and Space Administration. 
S.B. wishes to acknowledge her status as a settler on the ancestral lands of the Gabrielino/Tongva people. 

\end{acknowledgements}

\software{\texttt{tinygp} (github.com/dfm/tinygp), \texttt{jax} \citep{jax2018github}, \texttt{numpy} \citep{harris2020array}, \texttt{pandas} \citep{pandas}, \texttt{matplotlib} \citep{Hunter:2007aa}, \texttt{spectres} \citep{Carnall:2017a}, \texttt{species} \citep{Stolker:2020a}, \texttt{corner} \citep{corner}, \texttt{scipy} \citep{scipy}, \texttt{astropy} (\citealt{astropy:2013}, \citealt{astropy:2018}, \citealt{astropy:2022}), \texttt{orbitize!} \citep{Blunt:2020a}.}

\bibliography{HIP65426}
\bibliographystyle{aasjournal}
\bibliographystyle{yahapj.bst}

\appendix 

Here, we further evaluate the differences between orbit fits \#4, 6, and 7, which apply different priors on eccentricity, using a model selection lens. There exist many model selection metrics a statistician can use to pick an ``optimal'' model. In this section, we briefly motivate the definitions of two such metrics, the Akaike Information Criterion (AIC) and Watanabe-Akaike Information Criterion  (WAIC), drawing heavily from \citet{Gelman:2014a}, and use them to provide a different perspective on the explanation in Section \ref{sec:ecc_interp}.

We can choose to define a model's ``goodness'' by its ability to predict unseen data points. A ``good'' model will predict new, unseen data better than a ``bad'' model. However, a posterior comprises many models, each with its own probability based on existing data, and Bayesian model selection seeks to compare two \textit{posteriors}. In addition, we do not know a priori what we expect unseen data to look like; this is the purpose of model fitting. The logic \citet{Gelman:2014a} review is that one can define the ``expected (log) predictive density'' (elpd) for a new data point as the expected posterior probability of the new datapoint, weighted by the probability of the new datapoint itself (typically unknown):

\begin{equation}
    \mathrm{elpd} = \int{\log{p_{\rm post}(y_i)} f(y_i) dy}
\end{equation} where $p_{\rm post}$ is the posterior probability, $y_i$ is an unseen data point, and $f(y)$ is the unknown physical function generating new data. Taking it one step further, we can aim to compute the expected log \textit{pointwise} predictive density (elppd), which sums the elpd for each new datapoint over an arbitrary-sized unseen dataset. The problem then becomes determining an estimator of this elppd quantity. (See \citealt{Gelman:2014a} section 2.3 for more detail). 

Both the AIC and the WAIC take the approach of defining the elppd of an unseen dataset as the sum of two terms: the first term represents the model's ability to predict \textit{existing} datapoints, and the second term is an overfitting penalty. They differ in their definitions of each of these terms. Both of these estimators converge to the actual log pointwise predictive density (lppd)\footnote{The elppd is an estimator of a true underlying value, the lppd. The AIC and WAIC are two different definitions of elppd estimators.} under certain conditions.

The AIC computes the first term as the probability of the existing data \textit{given the maximum likelihood model}. The second term is simply the number of free parameters in the model. This definition has the pleasing property of being an unbiased estimator of the lppd for Gaussian posteriors which were computed using a flat prior. It may already be clear that we will choose to argue that the AIC is insufficient for the orbit described in this paper, as the orbital posteriors we seek to compare (e.g. Figure \ref{fig:corner}) are quite non-Gaussian. In addition, effects of prior volume are important for our posteriors, and the maximum likelihood estimate (which is the only relevant quantity for the AIC) is different from the MAP estimate for both eccentricity priors discussed in the previous section. Because the maximum likelihood estimate is approximately equal for the free-eccentricity and fixed-eccentricity models, while the number of free parameters differs, we can understand why the AIC metric favors the $e=0$ model (Table \ref{tab:model_selection}).

The WAIC uses the whole posterior, not just the maximum likelihood estimate, to compute both the first and second terms. Because of this, \cite{Gelman:2014a} call it a ``a more fully Bayesian approach for estimating the out-of-sample expectation.'' The first term is exactly the average predictive density for all existing datapoints (\citealt{Gelman:2014a} Eq 5), and the second term has a few definitions in the literature. In this article, we define the WAIC using \citet{Gelman:2014a} Eq 12, which sets the overfitting term to be the sum of the variances of the lppd values of individual existing datapoints. Under this definition, a model which is overfitting will have (on average) smaller variances in predicted posterior probabilities of existing datapoints. Because the WAIC includes information from the whole posterior, implicitly taking into account effects of prior volume, we might expect that the WAIC is a better estimator for our thoroughly non-Gaussian posteriors. Indeed, the WAIC shows a much less clear distinction between the various eccentricity models than the AIC (Table \ref{tab:model_selection}). The circular model is still preferred, but by a much smaller margin than the AIC suggests (and by an amount which many authors argue is ``essentially indistinguishable''). This motivates us to argue that, although the AIC prefers the fixed $e=0$ model, the three models are indistinguishable in terms of expected predictive ability. In other words, we do not rule out a moderate eccentricity, despite the AIC preference for a circular orbit, and determine that there is not enough evidence to conclusively select one of the three models as ``optimal.'' The preference for a low or moderate eccentricity over a high eccentricity ($\gtrsim0.8$), however, is driven by the lower likelihood at higher eccentricities, and is consistent across all three prior options.

\begin{deluxetable*}{cccc}
    \tablewidth{\textwidth}
    \tablecaption{Model comparison metrics for orbit-fits including all available astrometric data and varying the eccentricity prior. All values are computed relative to the $e=0$ model. The linear $e$ prior is given by Equation~(\ref{eq:ecc_prior}). For the AIC computation, M$_{\rm tot}$ and parallax were not included in the number of free parameters, since they were both highly constrained by their respective priors.
    \label{tab:model_selection}}
    \tablehead{ & Uniform $e$ prior & linear $e$ prior & $e=0$}
\startdata
$\Delta$AIC & +5.97 & +5.97 & 0.00 \\
$\Delta$WAIC & +0.22 & +0.04 & 0.00 \\
\enddata
\end{deluxetable*}

\end{document}